\documentclass[12pt]{article}

\usepackage[top=80pt,bottom=85pt,left=58pt,right=58pt]{geometry}
\usepackage{amssymb,amsmath,xypic}
\usepackage{slashed}
\usepackage{graphicx,color,xcolor,subfigure}
\usepackage{float}
\usepackage{sectsty}
\usepackage{cite}
\usepackage[debug,pageanchor=false]{hyperref}
\definecolor{link}{rgb}{.8,.15,.1}
\hypersetup{colorlinks=true,linkcolor=link,citecolor=link,urlcolor=link,linktocpage}

\makeatletter
\@addtoreset{equation}{section}
\makeatother

\renewcommand{\theequation}{\thesection.\arabic{equation}}

\newcommand{\mrm}[1]{\mbox{$\mathrm{#1}$}}
\newcommand{\beq}{\begin{equation}}
\newcommand{\eeq}{\end{equation}}
\newcommand{\bea}{\begin{eqnarray}}
\newcommand{\eea}{\end{eqnarray}}
\newcommand{\nn}{\nonumber}
\def\red{\textcolor[rgb]{0.98,0.00,0.00}}

\begin{document}

\begin{titlepage}

\begin{center}

\vskip .5in 
\noindent

{\Large \bf{Holographic $\frac12$-BPS surface defects in ABJM}}

\bigskip\medskip

Yolanda Lozano$^{a,b}$\footnote{ylozano@uniovi.es},  Niall T. Macpherson$^{a,b}$\footnote{macphersonniall@uniovi.es}, Nicol\`o Petri$^{c}$\footnote{petri@post.bgu.ac.il}, Anayeli Ram\'irez$^{d}$\footnote{Anayeli.Ramirez@mib.infn.it}  \\

\bigskip\bigskip\bigskip
{\small 

a: Department of Physics, University of Oviedo,
Avda. Federico Garcia Lorca s/n, 33007 Oviedo}

\medskip
{\small and}

\medskip
{\small 

b: Instituto Universitario de Ciencias y Tecnolog\'ias Espaciales de Asturias (ICTEA),\\
Calle de la Independencia 13, 33004 Oviedo, Spain}

\bigskip\medskip
{\small 

c: Department of Physics, Ben-Gurion University of the Negev, Be'er-Sheva 84105, Israel }

\bigskip\medskip
{\small 

d: Dipartimento di Fisica, Universit\`a degli studi di  Milano--Bicocca,\\and INFN, Sezione di Milano--Bicocca,\\ Piazza della Scienza 3, 20126 Milano, Italy }

\vskip 2cm 

     	{\bf Abstract }
     	\end{center}
     	\noindent
	
We study the class of $\text{AdS}_3\times \mathbb{CP}^3$ solutions to massive Type IIA supergravity with $\mathfrak{osp}(6|2)$ superconformal algebra recently constructed in \cite{Macpherson:2023cbl}. These solutions are foliations over an interval preserving $\mathcal{N}=(0,6)$ supersymmetry in two dimensions, that in the massless limit can be mapped to the $\text{AdS}_4\times \mathbb{CP}^3$ solution of ABJM/ABJ. We show that in the massive case extra NS5-D8 branes, that we interpret as $\frac12$-BPS surface defect branes within the ABJ theory, backreact in the geometry and turn one of the 3d field theory directions onto an energy scale, generating a flow towards a 2d CFT. We construct explicit quiver field theories that we propose flow in the IR to the $(0,6)$ SCFTs dual to the solutions. Finally, we show that the $\text{AdS}_3$ solutions realise geometrically, in terms of large gauge transformations, an extension to the massive case of Seiberg duality in ABJ theories proposed in the literature.
\noindent

\vfill
\eject

\end{titlepage}

\setcounter{footnote}{0}

\tableofcontents

\setcounter{footnote}{0}
\renewcommand{\theequation}{{\rm\thesection.\arabic{equation}}}

\section{Introduction}
		
The study of the AdS/CFT correspondence in low dimensions has been the subject of intense research in the last years,  particularly regarding the classification of new solutions. This study is prompted by the key role played by $\text{AdS}_3$ and $\text{AdS}_2$ spaces as near horizon geometries of extremal black strings/holes, and the unique possibility the AdS/CFT correspondence offers for their microscopical description. More recently, low dimensional holography has also found fruitful applications in the description of conformal defects in higher dimensional CFTs. These defects are typically realised as operator insertions that reduce the conformality and the number of supersymmetries of the higher dimensional theory where they are embedded. When the number of defects is large they backreact in the geometry and if some conformality is still preserved give rise to lower dimensional AdS geometries. The AdS/CFT correspondence formulated in these lower dimensional spaces offers a useful  framework to study these defects, especially when they are strongly interacting. 

 The variety of low dimensional AdS backgrounds is especially rich, given the many possibilities for the geometry of their internal spaces and the large variety of superconformal algebras that exist in low dimensions. Many new low dimensional AdS solutions have been reported in the last years whose construction has profited from the development of powerful techniques involving bispinors and G-structures. In this paper we will concentrate on $\text{AdS}_3/\text{CFT}_2$ holography. In these dimensions new results have been reported in \cite{Tong:2014yna}-\cite{Lozano:2022ouq}, plus the already mentioned \cite{Macpherson:2023cbl}, that will be the focus of this paper.
 
 Remarkably, some of these classification efforts have been accompanied by interesting investigations of the dual CFTs. Notable examples where this has been possible are the $\text{AdS}_3/\text{CFT}_2$ pairs constructed in \cite{Tong:2014yna,Couzens:2017way,Couzens:2017nnr,Lozano:2019emq,Lozano:2019jza,Lozano:2019zvg,Lozano:2019ywa,Couzens:2019mkh,Lozano:2020bxo,Faedo:2020nol,Faedo:2020lyw,Couzens:2021veb,Lozano:2022ouq}, which provide explicit settings for the microscopical description of black strings. This has been addressed in \cite{Haghighat:2015ega,Couzens:2019wls,Couzens:2021veb}. Notably, most known pairs rely on $\mathcal{N}=(n,n)$ or $(0,n)$ with $n\leq 4$ supersymmetry.
 
 In this paper we will investigate  $\text{AdS}_3/\text{CFT}_2$ holography with $(0,6)$ supersymmetry, taking as our starting point the  new class of solutions with $\mathcal{N}=(0,6)$  supersymmetries constructed recently in \cite{Macpherson:2023cbl}. Therefore this work opens up the interesting new possibility of investigating the $\text{AdS}_3/\text{CFT}_2$ correspondence in less standard supersymmetric settings\footnote{See also \cite{Conti:2023rul} which points the way towards similar constructions in 1 dimension less}. A second class of solutions with $\mathcal{N}=(0,5)$ supersymmetries was also constructed in \cite{Macpherson:2023cbl}, and it would be interesting to further extend the analysis in this paper to that class. The solutions with $\mathcal{N}=(0,6)$  supersymmetries  involve a three-dimensional complex projective space in their internal geometry, so they share many properties with the ABJM/ABJ geometry \cite{Aharony:2008ug,Aharony:2008gk}. The connection with the ABJM/ABJ theories will indeed play a key role in the identification of their dual CFTs. As we will see, this class of solutions can be regarded as an extension of the ABJM/ABJ solution of massless Type IIA supergravity to the massive case. In this extension one of the external directions becomes an energy scale, generating a flow towards an $\text{AdS}_3$ space. Remarkably, the Seiberg duality proposed in \cite{Bergman:2010xd}, in a particular massive generalisation of the ABJM/ABJ theory, is explicitly realised in these solutions as a large gauge transformation, in the spirit of \cite{Benini:2007gx,Aharony:2009fc}. 
 
 The paper is organised as follows. In section \ref{geometries} we summarise the main properties of the solutions with $\mathcal{N}=(0,6)$ supersymmetry reported in \cite{Macpherson:2023cbl}, and focus on the construction of local solutions that can be glued together with D8-branes. In this section we elaborate on the massless limit, briefly addressed in  \cite{Macpherson:2023cbl}. In this limit a change of variables allows to relate the direction interpreted as an energy scale to the radius of $\text{AdS}_4$, and the solutions reduce to the $\text{AdS}_4\times \mathbb{CP}^3$ solution of Type IIA supergravity, dual to the ABJM/ABJ theory. We study in detail how the free energy of the ABJM/ABJ theory arises and review some aspects of the theory, such as the realisation of Seiberg duality \cite{Aharony:2008gk,Aharony:2009fc}, that will be useful later on  in our study of the massive case. 

In section \ref{solutionsIIB} we construct a new class of solutions to Type IIB supergravity by acting with Abelian T-duality along one of the directions of the $\mathbb{CP}^3$, preserving $\mathcal{N}=(0,4)$ supersymmetry. We show that they reduce in the massless limit to the $\text{AdS}_4$ solution to Type IIB supergravity with $\mathcal{N}=4$ supersymmetry T-dual to $\text{AdS}_4\times \mathbb{CP}^3$, that constitute the basis for the brane construction of the ABJM/ABJ theory. 

 In section \ref{field-theory}, in analogy with the $\text{AdS}_4$ case, we use the Type IIB solutions to propose the brane set-up where the 2d dual CFTs should be realised. We argue that this brane intersection preserves $\mathcal{N}=(0,3)$ supersymmetry in 2d, that, we propose, should be enhanced to $\mathcal{N}=(0,6)$ in the IR, in analogy with what happens in the ABJM/ABJ theory, albeit with half the number of supersymmetries. We show that extra NS5-D7 branes render the brane intersection of the ABJM/ABJ theory two dimensional and half-supersymmetric. These branes find in this way an interpretation as $\frac12$-BPS surface defect branes within the ABJM/ABJ theory. This connects our findings with those in 
 \cite{Fujita:2009kw}, where D4-D8 (D5-D7 in Type IIB) defect branes were introduced as probe branes in the ABJM theory, in order to realise edge states in the Fractional Quantum Hall Effect. Our study shows that similar defects can be introduced in a way that preserves half of the supersymmetries and a $\mathfrak{osp}(6|2)$ superconformal subalgebra. 
 
 Specifically, the brane intersection that we propose consists on a brane box model in which D3-branes are bounded between two types of NS5-branes and bound states of NS5 and D5 branes, rotated with respect to each other, displayed along two field theory directions. This brane box model can be viewed as a $\mathcal{N}=(0,4)$ brane box model of the type constructed in \cite{Hanany:2018hlz}, in which some of the NS5 and D5 branes have been rotated with respect to each other. We argue that the  rotation reduces the SO(4)$_R$ R-symmetry group of the brane box construction in \cite{Hanany:2018hlz} to SO(3)$_R$, and thereby the supersymmetry from $\mathcal{N}=(0,4)$ to $\mathcal{N}=(0,3)$, with the extra SO$(3)$ remaining as a global symmetry. We summarise the field content of the $\mathcal{N}=(0,4)$ brane box models studied in \cite{Hanany:2018hlz}, to which we add the effect of the rotation. This renders some of the scalars living in the hypermultiplets in the Hanany-Okazaki brane boxes massive. As a result of our study we are able to associate a 2d quiver field theory to our brane intersection. Interestingly, we show that these quiver 
field theories realise a ``massive'' extension of Seiberg duality in ABJ proposed in \cite{Bergman:2010xd}. This is realised in terms of a brane creation effect when NS5-branes are crossed along the direction that became an energy scale in the 2d theory. This links Seiberg duality to large gauge transformations, albeit in one dimension less, in the spirit of \cite{Benini:2007gx,Aharony:2009fc}. We end the section with the computation of the holographic central charge, which, as we will see, includes higher derivative corrections. This represents a fairly non-trivial prediction for the central charge of the dual $(0,6)$ SCFTs. Finally, we point out an interesting relation between the holographic central charge and a product of electric an magnetic charges computed from the solutions, in the spirit of \cite{Lozano:2020txg,Lozano:2020sae,Lozano:2021rmk}. These references generalised 
the proposal in  \cite{Hartman:2008dq}, showing that the central charge in the algebra of symmetry generators of $\text{AdS}_2$ with an electric field is proportional to the square of the electric field, to fully-fledged $\text{AdS}_2$ string theory set-ups. Our result shows that this proposal extends to $\text{AdS}$ backgrounds with other dimensionalities and supersymmetries, and therefore suggests a more general interpretation of the central charge in terms of products of electric and magnetic charges. We conclude in section \ref{conclusions} with a summary of our results and a discussion of the main open problems that will be interesting to further investigate.

\section{The class of ${\cal N}=(0,6)$ AdS$_3\times \mathbb{CP}^3$ solutions to massive IIA}\label{geometries}
In this section we explore the new class of solutions originally found in \cite{Macpherson:2023cbl}. Our main aim is to understand what possible global solutions can be constructed by gluing local solutions together with D8 branes. We begin by reviewing some results of \cite{Macpherson:2023cbl} that we will make use of later in this section.\\
~\\
In \cite{Macpherson:2023cbl} the local form of AdS$_3$ solutions in massive IIA preserving the superconformal algebra $\mathfrak{osp}(6|2)$  were derived. Their NS sector can be expressed as
\begin{align}
\label{eq:metric}
\frac{ds^2}{2\pi}&= \frac{|h|}{\sqrt{2 h h''-(h')^2}}ds^2(\text{AdS}_3)+\sqrt{2 h h''-(h')^2}\bigg[ \frac{1}{4 |h|}dr^2+ \frac{2}{|h''|}
ds^2(\mathbb{CP}^3)\bigg],\nn\\[2mm]
e^{-\Phi}&=\frac{(|h''|)^{\frac{3}{2}}}{2\sqrt{\pi}(2 h h''-(h')^2)^{\frac{1}{4}}},~~~~H_3=dB_2,~~~~B_2=4\pi \left( -(r-l)+\frac{h'}{h''}\right)J.
\end{align}
Where $J$ is the Kahler form on $\mathbb{CP}^3$ which has unit radius, $l$ is an integer (whose role is clarified below) and $h$ is a function of $r$. A convenient parametrisation of $\mathbb{CP}^3$ is as a foliation of  $T^{1,1}$  over an interval, with respect to this we have
\begin{align}
ds^2(\mathbb{CP}^3)&= d\xi^2+\frac{1}{4}\cos^2\xi ds^2(\text{S}^2_1)+\frac{1}{4}\sin^2\xi ds^2(\text{S}^2_2)+ \frac{1}{4}\sin^2\xi\cos^2\xi (d\psi+ \eta_1+ \eta_2)^2,~~~ d\eta_i=- \text{vol}(\text{S}^2_i),\nn\\[2mm]
J&= \frac{1}{4}\sin^2\xi \text{vol}(\text{S}^2_2)-\frac{1}{4}\cos^2\xi \text{vol}(\text{S}^2_1)-\frac{1}{2}\sin\xi \cos\xi d\xi\wedge (d\psi+ \eta_1+\eta_2), \label{paramCP3}
\end{align}
where $(\xi,\psi)$ have periods $(\frac{1}{2}\pi,4\pi)$ and the 2-spheres have unit radius.\\
~\\
The fluxes making up the RR sector can be compactly expressed in terms of the NS 2-form potential as
\begin{align}
F_0&=-\frac{1}{2\pi}h''',~~~~F_2=B_2 F_0+2 (h''-(r-l)h''')J,\nonumber\\[2mm]
F_4&=\pi d\left(h'+\frac{hh'h''}{2h h''- (h')^2}\right)\wedge\text{vol}(\text{AdS}_3)+B_2\wedge F_2-\frac12 B_2\wedge B_2 F_0\nonumber\\[2mm]
&-4\pi (2h'+(r-l)(-2h''+(r-l)h'''))J\wedge J\label{RRgeometry}.
\end{align}
In regular regions of a solution the Bianchi identities imply that $F_0$ is constant, we thus have the ODE
\beq\label{eq:definingPDE}
h'''=-2\pi F_0.
\eeq
 Formally this is very similar to the class of AdS$_7$ solutions found in  \cite{Apruzzi:2013yva}, as presented in  \cite{Cremonesi:2015bld}. Of the remaining Bianchi identities, clearly that of $H_3$ is implied while those of the  RR fields are implied when \eqref{eq:definingPDE} holds. To this it is useful to introduce the following magnetic Page fluxes
\beq \label{Bfield}
\hat f= e^{-B_2}\wedge f
\eeq
for $f=f_0+f_2+f_4+f_6$ where $F_6=-\star F_4$ and $f$ means the magnetic part of $F=\sum_{n=0}^5F_{2n}$. We find these take the form
\begin{align}
\hat f_2&=  2 \left(h''- (r-l) h'''\right)J,\nn\\[2mm]
\hat f_4&= - 4\pi \left(2h'+(r-l)((r-l)h'''-2h'')\right)J\wedge J,\nn\\[2mm]
\hat f_6&=  \frac{16\pi^2}{3}\left (6h-(r-l)(6h'+(r-l)((r-l) h'''-3h''))\right)J\wedge J\wedge J. \label{eq:pagefluxes}
\end{align}
In regular regions of a solution these must be closed for the Bianchi identities of \eqref{RRgeometry} to hold and vice-versa. It is a simple matter to show that
\beq
d\hat f_{2n}= -\frac{1}{2\pi} (4\pi)^n(r-l)^n h''''\frac{1}{n!} dr\wedge J^n,\label{eq:PDEflux}
\eeq
so the magnetic Page fluxes are indeed closed in regular parts of a solution by \eqref{eq:definingPDE}.\\
~\\
When $F_0$ is constant \eqref{eq:definingPDE} implies $h$  is an order 3 polynomial that depends on 4  parameters $c_{1,2,3,4}$. Depending on how these are tuned the domain of the interval $r$ can differ and distinct physical behaviours can be realised. Assuming $F_0\neq 0$ and that $r$ is bounded at one end, which one can assume to be at $r=0$ without loss of generality, in \cite{Macpherson:2023cbl}  3 singular yet physical boundary behaviours were identified: namely that associated to a D8/O8 system of world volume AdS$_3\times \mathbb{CP}^3$,  an O2 plane extended in AdS$_3$ and backtracked on a G$_2$ cone of base $\mathbb{CP}^3$  and the singularity associated to the $d=3$ KK monopole one gets when reducing  the embedding of the hyper-Kahler 8-manifolds constructed in \cite{Gauntlett:1997pk} into $d=11$ supergravity to Type IIA, albeit extended in AdS$_3$ rather than $\text{Mink}_3$. These objects are supersymmetric wherever one places them in the space. To realise them at $r=0$ one must tune
\begin{subequations}
\begin{align}
&\text{D8/O8}:~~~~h= c_1+ c_2 r^3,~~~~c_{1,2}\neq 0\label{eq:O8}\\[2mm]
&\text{O2}:~~~~h=c_1+c_2 r+ \frac{c_2^2}{4c_1}r^2+ c_3 r^3,~~~~c_{1,2,3}\neq 0\\[2mm]
&\text{Monopole}:~~~~~h= (c_1+c_2 r) r^2,~~~c_{1,2}\neq 0.\label{eq:monopole}
\end{align}
\end{subequations}
However the domain of $r$ for all of these local solutions is semi-infinite, with $r=0$ an infinite proper distance from $r=\infty$. When $F_0=0$ one can show that one always locally recovers  AdS$_4\times \mathbb{CP}^3$. We explore this massless case in some detail in  section \ref{ABJM}.\\
~~\\
As stressed in \cite{Macpherson:2023cbl} globally $F_0$ needs only be piece-wise constant. If one defines $r=r_0$ to be a locus where $F_0$ is discontinuous, then in the neighbourhood of $r_0$ one has
\beq
dF_0=  \Delta F_0\delta(r-r_0),
\eeq
indicating the presence of a stack of D8-branes of charge $2\pi \Delta F_0$, where $\Delta F_0$ is the difference  between the constant values $F_0$ takes for $r > r_0$ and $r < r_0$. Of course for this really to be true the NS sector should be consistent with this behaviour. A D8 brane gives rise to a comparatively mild singularity for which all components of the metric and the dilaton neither blow nor tend to zero. They should however both be continuous as one crosses a D8 brane, which one can show amounts to imposing the continuity of
\beq
(h,(h')^2,h'')\label{eq:continuity conditions}
\eeq
as one crosses its loci. $h'''$ should of course be discontinuous or there is no D8. The NS 2-form should also be continuous modulo large gauge transformations, which is something else one needs to impose.

Another point is that one cannot expect to arbitrarily place branes in some supersymmetric solution without breaking supersymmetry. However  such a D8 brane, with no world-volume flux turned on, has been shown to preserve the full supersymmetry of this background when placed at
\beq
r_0=l.
\eeq
Like wise the source corrected Bianchi identities of such an object are  implied by \eqref{eq:PDEflux} -   only $dF_0$ experiences such a correction, the remaining $d\hat f_{2n}$ terms vanish as  $(r-l)^n\delta(r-l) \to 0$ for $n=1,2,3$. So one is free to make $F_0$ discontinuous at $r=l$ and the result is a supersymmetric stack of D8 branes. Naively this sounds like one can only place a single D8 brane stack on the interior of $r$,  but one must appreciate that $l$ does not need to be fixed globally. Indeed one can view the difference between $B_2$ with $l=0$ and the generic $B_2$ as a large gauge transformation, provided $l$ is an integer. Thus one can place arbitrary numbers of D8 branes along the interior of $r$, provided they are placed at integer points along $r$ and accompanied by the appropriate number of large gauge transformations of $B_2$. \\
~\\
In \cite{Macpherson:2023cbl} the above facts where used to construct a simple globally bounded solution, by gluing 2 copies of \eqref{eq:O8} together with a single D8 brane stack. In section \ref{eq:globalD8s}  we shall perform a more general analysis of possible  global solutions, however before doing so it will be instructive to  first consider the massless case in the next section.

\vspace{0.5cm}
\noindent {\bf Central charge:}

\noindent A closed expression for the calculation of the holographic central charge applicable to the solutions was given in \cite{Macpherson:2014eza,Bea:2015fja}, extending the computation in  \cite{Klebanov:2007ws} to more general backgrounds. In this reference the holographic central charge is computed for a metric and dilaton
\begin{equation}\label{cc1}
ds^2_{10}=a(\zeta,\vec{\theta})(dx_{1,d}^2+b(\zeta)d\zeta^2)+g_{ij}(\zeta,\vec{\theta})d\theta^id\theta^j, \qquad \Phi=\Phi(\zeta,\vec{\theta}),
\end{equation}
from
\begin{equation}
c_{hol}=\frac{3 d^d}{G_N}\frac{b(\zeta)^{d/2}(\hat{H})^{\frac{2d+1}{2}}}{(\hat{H}')^d},
\end{equation}
where
\begin{equation}\label{cc2}
\hat{H}=\Bigl(\int d\vec{\theta} e^{-2\Phi}\sqrt{\text{det}[g_{ij}]a(\zeta,\vec{\theta})^d}\Bigr)^2.
\end{equation}
This gives the simple expression 
\begin{equation}\label{centralcharge}
	c_{hol}=\frac12 \int dr \;(2hh''-(h')^2)
\end{equation}
for the solutions in \eqref{eq:metric}.

\subsection{The massless case: The ABJM/ABJ solution} \label{ABJM}

As shown in \cite{Macpherson:2023cbl}  in the massless limit the ABJM/ABJ solution  \cite{Aharony:2008ug,Aharony:2008gk} is recovered. To see this we can parametrise
\begin{equation}\label{eq:AdS4h}
h(r)=Q_2-Q_4r+\frac12 Q_6 r^2,
\end{equation}
where $Q_{2,4,6}$ are constants whose significance will become clear shortly. Then one can easily check that the change of variables
\begin{equation}\label{sinh}
\sinh{\mu}=\frac{Q_6 \,r -Q_4}{\sqrt{2Q_2Q_6-Q_4^2}}
\end{equation}
gives rise to the ABJM/ABJ metric and dilaton
\begin{eqnarray}
&&ds^2=\frac{4\pi \sqrt{2Q_2Q_6-Q_4^2}}{Q_6}\,\Bigl(\frac14 ds^2(\text{AdS}_4)+ds^2(\mathbb{CP}^3)\Bigr),\label{param}\\
&& ds^2(\text{AdS}_4)=d\mu^2+\cosh^2{\mu} \,ds^2(\text{AdS}_3),\quad e^{-\Phi}=\frac{Q_6^{\frac32}}{2 \sqrt{\pi} (2Q_2Q_6-Q_4^2)^\frac14}. \label{ABJMdilaton}
\end{eqnarray} 
The radius of $\mathbb{CP}^3$ is thus given by 
\begin{equation}\label{radius}
L=\Bigl(\frac{32\pi^2}{Q_6^2}(Q_2Q_6-\frac12 Q_4^2)\Bigr)^{1/4}.
\end{equation}
Therefore there is just one local solution when $F_0=0$ and it is an AdS$_4$ vacuum preserving twice the supersymmetries of generic solutions within this class. This is actually the only regular solution also.

One can now use the Page fluxes in  \eqref{eq:pagefluxes} to compute the various brane charges, leading to
\begin{align}
\frac{1}{2\pi}\int_{\mathbb{CP}^1} \hat f_2= Q_6,~~~~
\frac{1}{(2\pi)^3} \int_{\mathbb{CP}^2}\hat f_4&=Q_4,~~~~\frac{1}{(2\pi)^5} \int_{\mathbb{CP}^3}\hat f_6= Q_2. \label{QsAdS4}
\end{align}
Thus the constants in \eqref{eq:AdS4h},  $Q_p$, are identified with the Page charges of $D_p$ branes, for $p=2,4,6$. In turn, in this case one finds $H_3=0$ and $\hat{f}_4=-f_2\wedge B_2$. The NS5 and the D4 branes are thus not physical, since they annihilate to nothing and only their fluxes remain \cite{Aharony:2008gk}. $B_2$ is given by
\begin{equation}\label{B2massless}
B_2=-4\pi \frac{Q_4}{Q_6} J,\qquad \text{such that} \qquad b=-\frac{Q_4}{Q_6},
\end{equation}
and the discrete holonomy in the construction of  \cite{Aharony:2008gk} arises.
As discussed there this fractional charge is associated to the fact that in the presence of $Q_6$ units of $f_2$ flux a NS5-brane wrapped on the $\mathbb{CP}^2$ must have $Q_4$ D4-branes wrapped on the $\mathbb{CP}^1$ ending on it.

Table \ref{tableABJM} summarises the brane set-up associated to the ABJM/ABJ solution in its Type IIB realisation \cite{Aharony:2008ug}\footnote{After a T-duality along the Hopf fibre of the S$^3$ included in the $T^{1,1}$ in the $\mathbb{CP}^3$ has been performed (see the next section).}. The D3-branes stretch on the $\psi$-circle, intersecting one NS5'-brane and one $(1,k)$ 5'-brane\footnote{As we recall below, the numbers of branes in the field theory do not actually coincide with the Page charges computed from the solution. This explains our different notation for the charges in this discussion.}, extended along  $(x^0,x^1,r)$ and the $[3,7]_\theta$, $[4,8]_\theta$, $[5,9]_\theta$ directions, with $\tan{\theta}=k$. On top of this there are $M$ fractional branes stretched just along one segment of the circle, which as mentioned above are not actual branes since they are unstable. The brane system preserves $\mathcal{N}=3$ supersymmetry in 3d, and this is enhanced to $\mathcal{N}=6$ in the IR \cite{Aharony:2008ug}.
\begin{table}[http!]
\renewcommand{\arraystretch}{1}
\begin{center}
\scalebox{1}[1]{
\begin{tabular}{c| c cc  c c  c  c c c c}
 branes & $x^0$ & $x^1$ & $r$ & $x^3$ & $x^4$ & $x^5$ & $\psi$ & $x^7$ & $x^8$ & $x^9$ \\
\hline \hline
$N\,\mrm{D}3$ & $\times$ & $\times$ & $\times$ & $-$ & $-$ & $-$ & $\times$ & $-$ & $-$ & $-$ \\
$\mrm{NS}5'$ & $\times$ & $\times$ & $\times$ & $\times$ & $\times$ & $\times$ & $-$ & $-$ & $-$ & $-$ \\
$(1,k) 5'$ & $\times$ & $\times$ & $\times$ & $\cos{\theta}$ & $\cos{\theta}$ & $\cos{\theta}$ & $-$ & $\sin{\theta}$ & $\sin{\theta}$ & $\sin{\theta}$ \\
\end{tabular}
}
\caption{Brane intersection describing the ABJM brane system. $(x^0,x^1,r)$ span the 3d field theory. The NS5$'$ brane and D5$'$ brane at $\psi=\pi$ are rotated the same angle on the $[3,7]$, $[4,8]$ and $[5,9]$ directions.
The brane intersection preserves 6 Poincar\'e supersymmetries ($\mathcal{N}=3$ in 3d).} \label{tableABJM}
\end{center}
\end{table}

In the massless limit a large gauge transformation $b\rightarrow b+1$ maps the theory onto itself, and therefore the associated change in the quantised charges, given by equations \eqref{changeQs} for $Q_8=0$,
\begin{eqnarray}
&&Q_2\rightarrow Q_2-Q_4+\frac12 Q_6, \nonumber\\
&&Q_4\rightarrow Q_4-Q_6, \label{Pagetransmassless}
\end{eqnarray}
must be a symmetry of the theory. Indeed, as shown in \cite{Aharony:2009fc}, these transformations generate Seiberg dualities relating the IR behaviour of  different 3d $\mathcal{N}=3$ Chern-Simons matter theories. In order to see this it is necessary to recall that due to the non-trivial topology of the $\mathbb{CP}^3$ the quantised charges emerging from the solutions do not coincide with the numbers of branes in the field theory. 
This happens because the fractional worldvolume field strength that is needed to cancel the Freed-Witten anomaly \cite{Freed:1999vc} associated to the $\mathbb{CP}^2$, being a non-spin manifold, induces charge on the branes wrapped on the $\mathbb{CP}^2$. On top of this there is a non-vanishing contribution of higher curvature terms for these branes \cite{Bergman:2009zh}. 
Adding these contributions the following linear relations between the quantised charges and the numbers of D2, D4 and D6 branes, that we have denoted by $N,M$ and $k$, respectively, are found  \cite{Aharony:2009fc}
\begin{eqnarray}
&&Q_2= N+\frac{k}{12}, \nonumber\\
&&Q_4=M-\frac{k}{2}, \nonumber\\
&&Q_6= k.\label{change}
\end{eqnarray}
Substituting these relations in \eqref{B2massless} the shift of $B_2$ by $2\pi J$ required to cancel the Freed-Witten anomaly naturally arises. Moreover, one can see that the transformations \eqref{Pagetransmassless} translate into
\begin{equation}\label{Seiberg}
N\rightarrow N+k-M, \qquad M\rightarrow M-k,
\end{equation}
for the field theory parameters. These transformations map  the 3d Chern-Simons matter theory with gauge group U$(N+M)_k\times$U$(N)_{-k}$ to the one with gauge group U$(N)_k\times$U$(N-M+k)_{-k}$, and successively. These are the Seiberg duality transformations that relate the IR behaviour of the different Chern-Simons matter theories found in \cite{Aharony:2008gk,Aharony:2009fc}. 
One can check that the central charge computed using the expression \eqref{centralcharge}, that resorts in the massless case to
\begin{equation}\label{centralcharge2d}
c_{hol}=\frac12 (2Q_2Q_6-Q_4^2)=Nk- \frac12 M (M-k)-\frac{1}{24}k^2,
\end{equation}
per unit $r$-interval, where we have used the relations \eqref{change} to write it in terms of the field theory charges,
is indeed invariant under the transformations \eqref{Pagetransmassless} and \eqref{Seiberg}. 
Note that in order to recover the right scaling of the central charge of the 3d theory we have to multiply the previous expression by the radius squared of $\text{AdS}_4$\footnote{Divided by $2\pi$, in order to match the different conventions used for the 2d and 3d metrics.}. This yields
\begin{equation}
c_{hol}^{(3d)}=\frac{(2Q_2Q_6-Q_4^2)^{3/2}}{Q_6}=\frac{L^6 Q_6^2}{64\pi^3}=\frac{1}{k}\Bigl(2Nk-M(M-k)-\frac{1}{12}k^2\Bigr)^{3/2}, \label{cholmassless}
\end{equation}
using again \eqref{change}.
This matches the result obtained
for the ABJM solution using equations \eqref{cc1}-\eqref{cc2} for $d=2$ and the metric and dilaton given by \eqref{param},\eqref{ABJMdilaton}. This expression is  interpreted as the free energy of the 3d CFT in the supergravity approximation. However, given that we have used the shifted charges given by \eqref{change} we obtain the planar free energy (at large 't Hooft parameter) with higher derivative sub-leading corrections  \cite{Bergman:2009zh,Bergman:2013qoa}. As in these references we can use the Maxwell charge for the D2-branes, given by,
\begin{equation}\label{Maxwell}
Q_{2}^M=Q_{2}+b\,Q_{4}+\frac12 b^2\, Q_{6}=N-\frac{1}{24}k-\frac12\frac{M^2}{k}+\frac12 M,
\end{equation}
to write
\begin{equation}
c_{hol}^{(3d)}=2\sqrt{2k}(Q_{2}^M)^{3/2}=2^{3/2}\,k^2\,{\hat{\lambda}}^{3/2} ,
\end{equation}
with $\hat{\lambda}=\frac{Q_{2}^M}{k}$ the shifted 't Hooft coupling parameter found in \cite{Drukker:2010nc}. This expression was shown to agree with the strong coupling expansion of the free energy computed from the matrix model calculation of the path integral on S$^3$ \cite{Drukker:2010nc,Herzog:2010hf,Fuji:2011km,Marino:2011eh}. Note that in three dimensions the useful role played by conformal anomaly coefficients in even dimensional CFTs in measuring the number of degrees of freedom is now played by matrix integral calculations of the path integral, using supersymmetric localization methods \cite{Kapustin:2009kz}.

\subsection{Analysis of global solutions with D8 branes}\label{eq:globalD8s}
In this section we perform an analysis of the global solutions it is possible to construct by placing D8-branes along the interior of $r$. Our aim is not to be completely exhaustive but rather to shed light on the wide variety of possibilities, and assemble the necessary building blocks for their construction.\\
~\\
In order to construct broad classes of solutions we shall segment the interval spanned by $r$ into intervals of unit length with $h$ taking the form 
\begin{equation}\label{hprofile}
h_l(r)=Q_2^l-Q_4^l(r-l)+\frac12 Q_6^l (r-l)^2-\frac16 Q_8^l (r-l)^3,
\end{equation}
in the interval $r\in [l,l+1]$ for $Q_2,...Q_8$ constants. We shall assume that in the $l$'th interval 
\beq
B_2=4\pi \left( -(r-l)+\frac{h_l'}{h_l''}\right)J,
\eeq
so that when  $Q_8^l\neq Q_8^{l-1}$, the resulting stack of D8 branes at $r=l$ is supersymmetric. Consistency requires that as we cross between the 2 adjacent intervals at $r=l$ we have that
\begin{equation}
b=\frac{1}{4\pi^2}\int_{\mathbb{CP}^1}B_2,
\end{equation}
is such that $b\rightarrow b+z$ for $z\in\mathbb{Z}$, so that a large gauge transformation is performed creating $z$ units of NS5 brane charge. Again the constants appearing in $h_l$,  $Q_2^l$, $Q_4^l$, $Q_6^l$ and $Q_8^l$ are the Page charges associated to  D2, D4, D6 and D8 branes in each interval, ie using \eqref{RRgeometry} one finds
\begin{align}
2\pi F_0&= Q_8^l,~~~~\frac{1}{2\pi}\int_{\mathbb{CP}^1} \hat f_2= Q_6^l,\nn\\[2mm]
\frac{1}{(2\pi)^3} \int_{\mathbb{CP}^2}\hat f_4&=Q_4^l,~~~~\frac{1}{(2\pi)^5} \int_{\mathbb{CP}^3}\hat f_6= Q_2^l, \label{Qs}
\end{align}
in the $l$'th cell.\\
~\\
\noindent {\bf Metric reality:}\\[2mm]
The first  thing we need to ensure is that the warp factors of the metric are real in each cell. The metric is real and positive when $-(h')^2+2h'' h\geq 0$, with equality only possible at the upper or lower bound of $r$ in a global solution. In a generic unit cell one should have $-(h')^2+2h'' h> 0$ which implies $h h''>0$, in what follows we shall assume this is solved as
\beq
h>0,~~~~h''>0,\label{eq:hassumption}
\eeq
without loss of generality. It is possible to show that $-(h')^2+2h'' h> 0$ across the $l$'th cell if the inequality holds at  $r=l$ and $r=l+1$. This amounts to imposing one of two sets of constraints on the charges
\begin{subequations}
\begin{align}
&Q^l_2\geq 0,~~~~Q^l_6\geq 0,~~~~Q^l_8\leq 0,~~~~2Q^l_2 Q^l_6\geq (Q^l_4)^2,\label{eq:simplesol}\\[2mm]
&Q^l_2\geq 0,~~~~Q^l_6> Q^l_8\geq 0,~~~~2(Q^l_6-Q^l_8)(Q^l_2+\frac{1}{2}Q^l_6-Q^l_4-\frac{1}{6}Q^l_8)\geq (Q^l_4-Q^l_6+\frac{Q^l_8}{2})^2,\label{eq:simplesol2}
\end{align}
\end{subequations}
in generic cells, which both leave the sign of $Q_4$ arbitrary.\\
~\\
\noindent {\bf Metric continuity:}\\[2mm]
Next we need to impose the continuity conditions \eqref{eq:continuity conditions} as we cross each D8 brane at the intersections of the cells.
To impose the continuity constraints we must distinguish between two cases
\beq
h'_{l-1}(l)= \pm h'_l(l).
\eeq
Let us first assume $h'_{l-1}(l)=  h'_l(l)$; the continuity conditions for $h_l$, $h'_l$, $h''_l$ across intervals force the conditions
\begin{eqnarray}
&&Q_2^l= Q_2^{l-1}-Q_4^{l-1}+\frac12 Q_6^{l-1}-\frac16 Q_8^{l-1},\nonumber \\
&&Q_4^{l}=Q_4^{l-1}-Q_6^{l-1}+\frac12 Q_8^{l-1}, \nonumber\\
&&Q_6^l= Q_6^{l-1}- Q_8^{l-1}, \label{changeQs}
\end{eqnarray}
between the $(l-1)^{th}$ and $l^{th}$ cells. We note that if \eqref{eq:simplesol}  holds in the  $(l-1)^{th}$ cell the continuity conditions impose that it also holds in the $l^{th}$ cell provided that $Q^l_8<0$. Conversely if \eqref{eq:simplesol2} holds in the $(l+1)^{th}$ cell it is guaranteed to also hold in the $l^{th}$ if $Q^l_8>0$. We thus see if one imposes $\eqref{eq:simplesol}$ in a single cell it is guaranteed to also hold in each subsequent cell to the right of it satisfying $Q^l_8\leq 0$. An analogous statement also holds for cells satisfying $Q^l_8\geq 0$ to the left of a cell satisfying \eqref{eq:simplesol2}.

Using that
\begin{equation} \label{NS5creation}
B_2^l=4\pi \frac{-Q_4^l+\frac12 Q_8^l (r-l)^2}{Q_6^l-Q_8^l (r-l)}\, J,
\end{equation}
in the $[l,l+1]$ interval, one can check that $b^{l}(l)=b^{l-1}(l)+1$, meaning that the shift in $B_2$ is indeed a large gauge transformation as we cross the D8 at $r=l$, and so 1 unit of NS5-brane charge is created as we do so. Note that the number of NS5-branes at $r=l$ is still fixed by the numbers of D4 and D6 branes in such interval. Thus, the D4-branes keep being fractional branes as in the massless case.

We can use the relation \eqref{changeQs} to express the charges in the $l$'th cell in terms of those in some other $\tilde{l}$'th cell, with $\tilde{l}< l$; assuming $h'_{l-1}(l)=  h'_l(l)$ is satisfied across all D8 branes,  we find
\begin{eqnarray}
&&Q_2^l= Q_2^{\tilde{l}}-\Delta l Q_4^{\tilde{l}}+\frac12 (\Delta l)^2Q_6^{\tilde{l}}-\frac16 \sum_{i=1}^{\Delta l}(1-3i+3i^2 )Q_8^{l-i},\nonumber \\
&&Q_4^{l}=Q_4^{\tilde{l}}-\Delta l Q_6^{\tilde{l}}+\frac12\sum_{i=1}^{\Delta l}(2i-1) Q_8^{l-i}, \nonumber\\
&&Q_6^l= Q_6^{\tilde{l}}-\sum_{i=1}^{\Delta l} Q_8^{l-i}, \label{sumQP}
\end{eqnarray}
where $\Delta l= l-\tilde{l}$.

If we instead assume $h'_{l-1}(l)= - h'_l(l)$ things are a bit more complicated. The continuity conditions impose the same constraints on $(Q_2^l,Q_6^l)$ as in \eqref{changeQs} but that of $Q_4^l$ is modified as 
\beq
Q_4^{l}=-(Q_4^{l-1}-Q_6^{l-1}+\frac12 Q_8^{l-1}),\label{eq:signchangerule}
\eeq
which now implies that 
\beq
b^{l}(l)-b^{l-1}(l)=\frac{2Q_4^{l-1}-Q_6^{l-1}}{Q_6^{l-1}-Q_8^{l-1}}.
\eeq
This is by no means automatically a large gauge transformation, for that we must impose that the LHS of the above is an integer, which has consequences for the charges in both the $l$'th and $(l-1)$'th cells, namely
\begin{align}
2Q_4^l&=-(z-1)Q_6^l,\nn \\[2mm]
2Q_4^{l-1}&=(z+1)Q_6^{l-1}-z Q_8^{l-1},
\end{align}
for $z\in \mathbb{Z}$ - note that these are not independent, they imply each other given \eqref{eq:signchangerule}. We again find that if \eqref{eq:simplesol} holds in the $(l-1)$'th cell it also holds in the $l$'th cell when $Q_8^{l}\leq 0$ and that if  \eqref{eq:simplesol2} holds in the $l$'th then it also holds in the $(l-1)$'th cell when $Q_8^{l-1}\geq 0$.

We can again sum the recursion relations, this time assuming that $h'_{l-1}(l)=  h'_l(l)$ is satisfied across all D8s except the one at $r=l$. We find
\begin{eqnarray}
&&Q_2^l= Q_2^{\tilde{l}}-\Delta l Q_4^{\tilde{l}}+\frac12 (\Delta l)^2Q_6^{\tilde{l}}-\frac16 \sum_{i=1}^{\Delta l}(1-3i+3i^2 )Q_8^{l-i},\nonumber \\
&&Q_4^{l}=\frac{1}{2}(1-z)(Q_6^{\tilde{l}}-\sum_{i=1}^{\Delta l}Q^{l-i}_8), \nonumber\\
&&Q_6^l= Q_6^{\tilde{l}}-\sum_{i=1}^{\Delta l} Q_8^{l-i}, \label{sumQM}
\end{eqnarray}
which is subject to the additional constraint
\beq
2 Q_4^{\tilde{l}}-(2\Delta l+z-1)Q_6^{\tilde{l}}+ \sum_{i=2}^{\Delta l}(2(i-1)+z)Q_8^{l-i}+z Q^{l-1}_8=0,
\eeq
which one can take to define, for instance, $Q^{l-1}_8$.
The most general configuration one can construct will have  $h'_{l-1}(l)= - h'_l(l)$  at an arbitrary number of cell intersections, with each such intersections being  separated by an arbitrary number of cells whose intersections obey $h'_{l-1}(l)= + h'_l(l)$ - clearly this leads to infinite possibilities. One can always compute the charges in a given cell of an arbitrary  solution by combining the results of \eqref{sumQP} and \eqref{sumQM}.\\
~\\
\noindent {\bf Simple global solutions:}\\[2mm]
To construct global solutions we need to decide how the space is going to start and end: If we choose to bound the space from below we can do so without loss of generality in the $0$'th cell at $r=0$, we must choose one of the profiles in \eqref{eq:O8}-\eqref{eq:monopole} so one achieves this with a physical singularity. Each of these choices  amounts to tuning the charges in the $0$'th cell as
\begin{align}
&\text{D8/O8 at } r=0:~~~~~~~~Q_4^0= Q_6^0=0,\label{eq:starspace}\\[2mm]
&\text{O2 at } r=0:~~~~~~~~~~~~2Q_2^0Q_6^0= (Q_4^0)^2\nn,\\[2mm]
&\text{Monopole at } r=0:~~~~ Q_2^0=Q^0_4=0 ,\nn
\end{align}
where in each case $Q_8^0<0$ and $Q_6^0,Q_2^0\geq 0$ ensures a well defined metric throughout the first cell, and so also in every subsequent cell for which $Q^l_8< 0$.  As fixing $F_0=0$ in our class makes it locally AdS$_4\times \mathbb{CP}^3$, as explored earlier, clearly we can also choose to begin the space with an AdS$_4$ boundary at  $r=-\infty$ by simply fixing $Q^l_8=0$ in every cell below some lower bound $l=l_0$ were we begin placing D8 brane sources. This requires
\beq
\text{AdS}_4~\text{boundary at: } r=-\infty:~~~~~~~ Q^l_8=0,~~~ \text{for}~~~  l<l_0,
\eeq  
one also needs to arrange for $2Q_2^lQ_6^l> (Q_4^l)^2$ below $r=l_0$, but then provided that we keep $Q^l_8\leq 0$ for $l\geq l_0$ we are then guaranteed to have a well defined metric.

Each of the previous options allows us to continue the space arbitrarily towards $r=+\infty$, placing D8 branes at arbitrary integer values of $r$ as we go as long as Q$^l_8\leq 0$ in each cell. At some point though we need to arrange for the space to terminate physically, one option is to keep  $Q^l_8< 0$ for $l=0,...,P-1$ and place a final D8 brane at $r=P$. In this scenario by far the easiest physical option to end the space is to have an AdS$_4$ boundary at $r=+\infty$, as all this requires one to do is fix
\beq
\text{AdS}_4~\text{boundary at: } r=+\infty,~~~~~~~ Q^l_8=0,~~~ l\geq P,
\eeq  
but this is not the only option at our disposal. We can instead choose to terminate the space in the $P$'th cell at $r=P+1$, for which we must again select a profile as in \eqref{eq:O8}-\eqref{eq:monopole}, but such that the singularity lies in the appropriate place, which amounts to taking $h$ in the final cell to be of the form
\beq
h=Q^{P+1}_2- Q^{P+1}_4 (r-P-1)+ \frac{1}{2}Q^{P+1}_6(r-P-1)^2- \frac{1}{3!}Q^{P}_8 (r-P-1)^3,\label{eq:upper}
\eeq
where $Q^{P+1}_{2,4,6}$ are the Page charges in the would be $(P+1)$'th cell, related to those in $P$'th cell by \eqref{changeQs}.
The distinct boundary behaviours are given by tuning
\begin{align}
&\text{D8/O8 at } r=P+1:~~~~~~~~Q^{P+1}_4= Q^{P+1}_6=0,\label{eq:end space}\\[2mm]
&\text{O2 at } r=P+1:~~~~~~~~~~~~2Q^{P+1}_2 Q^{P+1}_6= (Q^{P+1}_4)^2\nn,\\[2mm]
&\text{Monopole at } r=P+1:~~~~ Q^{P+1}_2=Q^{P+1}_4=0 ,\nn
\end{align}
where we still require that $Q^{P+1}_2,Q^{P+1}_6>0$ as before, but now $Q_8^P>0$ in each case. Notice that these imply the conditions \eqref{eq:simplesol2}, so one can equally well work backwards from $r=P+1$, gluing on cells with  $Q_8^l>0$ and have guaranteed metric positivity when the continuity constraints hold - we shall return to this point shortly.  Under the assumption that $h'$ does not change sign as we cross the D8 at $r=P$, it does not take long to establish that terminating at a D8/O8 is impossible, ie one finds
\beq
2Q^{P+1}_2+\frac{1}{3}(1+3P+6 P^2)Q_6^0= \sum_{i=1}^P(2P -i)i Q_8^{P-i},
\eeq
which is impossible to solve as the LHS is positive definite while the RHS is negative definite. Ending with a monopole under these assumptions is possible, but only when $Q^0_4 > 0$, we find we must tune the charges as
\begin{align}
Q_4^0&=(P+1)Q_6^0-\frac{1}{2}Q_8^P-\frac{1}{2}\sum_{i=1}^P(2i+1)Q_8^{P-i},\nn\\[2mm]
6Q_2^0&=3(P+1)^2Q_6^0-(2+3P)\sum_{i=0}^P Q^i_8+\sum_{i=1}^Pi (i-(2P+1))Q_8^{P-i},\nn\\[2mm]
Q_6^0&>\sum_{i=0}^P Q^i_8= Q^P_8+\sum_{i=1}^PQ^{i-1}_8,
\end{align}
where $Q_4^0>0$ follows from these and the fact that we must have $\tilde{Q}^P_6>0$. It is also possible to end the space with a O2 plane in this fashion, but solving $2\tilde{Q}_2^P\tilde{Q}_6^P= (\tilde{Q}_4^P)^2$ given \eqref{sumQP} leads to rather complicated expressions we will omit. We conclude that if $Q^l_8\leq 0$ for $l=0,...,P-1$ and $Q^P_8\neq 0$ then when $h'$ does not change sign at $r=P$ the possible global solutions are
\beq
(\text{AdS}_4,~\text{O2})\to(\text{AdS}_4,~ \text{O2},~ \text{Monopole}),
\eeq
with an arbitrary number of D8 branes on the interior of the $r$ interval.

If  $h'$ does change sign at $r=P$ it becomes possible to terminate the space with a D8/O8 singularity, this imposes
\begin{align}
Q^0_6&=\sum_{i=0}^PQ_8^i\geq 0,~~~Q_2^0\geq P Q^0_4+\frac{1-3P^2}{6}\sum_{i=1}^PQ_8^i+ 3\sum_{i=1}^Pi(i-1) Q_8^{i-1} ,\nn\\[2mm]
Q^0_4&=-Q_8^P+\frac{1}{2}(1+2P)\sum_{i=1}^P Q^i_8-\sum_{i=1}^Pi Q^{P-1}_8,~~~~z=0
\end{align}
which is compatible with all physical options for the lower bound of the space except the monopole.  It is also impossible to end the space with a monopole when $h'$ changes sign at $r=P$ as the necessary positivity constraints on the charges cannot be satisfied. Finally, although we will again not give the details due to their complexity, it is possible to end the space with an O2 in this fashion. We thus find that when $Q^l_8\leq 0$ for $l=0,...,P-1$ and $Q^P_8\neq 0$, then if  $h'$ does  change sign at $r=P$ 
\beq
(\text{D8/O8},~\text{AdS}_4,~\text{O2})\to(\text{D8/O8},~\text{AdS}_4,~ \text{O2})
\eeq
are the possible global solutions.\\
~\\
\noindent {\bf Symmetric profiles:}\\[2mm]
In the previous examples of global solutions we only allowed for one cell with $Q^l_8>0$. To end this section we will  consider the more symmetric case of an even number of positive and negative charges. 

One can begin the space at $r=0$ with any of the behaviours in \eqref{eq:starspace}, or arrange $Q^0_8$ to be the first non zero values of $Q^l_8$ such that we have an AdS$_4$ boundary at $r=-\infty$. Either way, we continue the space in a well defined manor up to $r=P$ so long as $Q^l_8<0$ for $l=0,...,P-1$ and we impose \eqref{eq:simplesol} in the $0$'th cell.  The idea is to then essentially glue the profile for $h$ to itself in the $P$'th cell: To this end it is useful to fix the charges in the $P$'th cell as
\begin{align}
Q^P_2&=Q^{P-1}_2-Q^{P-1}_4+\frac{1}{2}Q^{P-1}_6-\frac{1}{3!}Q^{P-1}_8,~~~~Q^{P}_8=-Q^{P-1}_8\nn\\[2mm]
Q^P_4&=-Q^{P-1}_4+Q^{P-1}_6-\frac{1}{2}Q^{P-1}_8,~~~~Q_6^{P}=Q^{P-1}_6-Q^{P-1}_8,\label{eq:chargessym}
\end{align}
which satisfies the continuity conditions at $r=P$ with $h'$ changing sign at this point. This tuning has the benefit of automatically solving the metric reality constraints: As $Q^{P}_8\geq 0$ the appropriate constraints are \eqref{eq:simplesol2}, which under \eqref{eq:chargessym} become
\begin{align}
&Q_8^{P-1}\leq 0,~~~~Q^{P-1}_6\geq 0,~~~ Q^{P-1}_2\geq 0,~~~~ 2Q_2^{P-1}Q_6^{P-1}\geq (Q^{P-1}_4)^2,\nn\\[2mm]
&Q_2^{P-1}-Q_4^{P-1}+\frac{1}{2}Q_6^{P-1}-\frac{1}{3!}Q^{P-1}_8\geq0,
\end{align}
the first line of which is just the \eqref{eq:simplesol} version of the metric reality conditions for the $(P-1)$'th cell, while the second line is implied by these. If we now consider the metric continuity constraints \eqref{changeQs} (ie without $h'$ changing sign) at $r=P+1$ we find that under \eqref{eq:chargessym} they are mapped to 
\beq
Q^{P+1}_{2}=Q^{P-1}_{2},~~~~~Q^{P+1}_{4}=-Q^{P-1}_{4},~~~~~Q^{P+1}_{6}=Q^{P-1}_{6}
\eeq
If we further identify $Q^{P+1}_{8}=-Q^{P-2}_{8}$ then the metric reality constraints of the $(P+1)$'th cell are implied by those of the $(P-2)$'th cell. We can continue in this fashion to construct a well defined global solution by simply identifying the charges in the $(P+n)$'th cell as 
\beq
Q^{P+n}_{2}=Q^{P-n}_{2},~~~~~Q^{P+n}_{4}=-Q^{P-n}_{4},~~~~~Q^{P+n}_{6}=Q^{P-n}_{6},~~~~~Q^{P+n}_{8}=-Q^{P-n-1}_{8}.\label{eq:symmprofile},
\eeq
for $n=1,...,P-1$. In such symmetric profiles $h'$ changes sign at the D8 at $r=P$, and only at this D8. We thus also need to ensure that the transformation of $B_2$ across $r=P$ is a large gauge transformation, which amounts to imposing
\beq
2 Q_4^{0}-(2(P-1)+z-1)Q_6^{0}+ \sum_{i=2}^{P-1}(2(i-1)+z)Q_8^{P-i}+z Q^{P-1}_8=0,\label{eq:constraintsss}
\eeq
which we can take as the definition of $Q_8^{P-1}$, which one can always ensure is negative by suitably choosing the integer $z$. In the $(2P-1)$'th cell we have $n=P-1$ and by using \eqref{changeQs}, the profile in this cell can then be written as
\beq
h=Q^0_2-(- Q^0_4) (r-2P)+ \frac{1}{2}Q^0_6(r-2P)^2- \frac{1}{3!}(-Q^0_8) (r-2P)^3
\eeq
making clear that if we begin the space at $r=0$ with one of the behaviours in \eqref{eq:starspace}, then it ends at $r=2P$ in the same fashion - likewise an  AdS$_4$ lower boundary will interpolate to an upper boundary AdS$_4$. It is thus possible to construct global solutions interpolating between the following boundary behaviours in this way
\beq
\text{D8/O8}\to \text{D8/O8},~~~~\text{Monopole}\to \text{Monopole},~~~~\text{O2}\to \text{O2},~~~~\text{AdS}_4\to \text{AdS}_4.
\eeq
It is also not hard to further generalise - for instance one can separate the negative and positive $Q_8^l$ by a region  with $Q^l_8=0$ of arbitrary length in a similar fashion, or one could take a portion of a symmetric profile then glue it to a local AdS$_4$ region. Let us stress that there is much more that one can do than we have discussed explicitly, however this section does provide the necessary building blocks to construct all these possibilities.\\
~\\
\noindent {\bf Central charge:}\\[2mm]
Finally, let us comment on the central charge of generic solutions: The holographic central charge \eqref{centralcharge} clearly depends on how we start and end the space, but we can make some general comments: The expression \eqref{centralcharge} can be compartmented into its contribution in each cell, with $h_l$ given by equation \eqref{hprofile} one finds the contribution between $r=l$ and $r=l+1$  to be 
\beq\label{centralchargeexplicit}
c^{l}_{hol}=\frac12 \int_{l}^{l+1}dr \Bigl(2h_lh_l''-(h_l')^2\Bigr)=2Q_2^lQ_6^l-(Q_4^l)^2-Q_8^l\,(Q_2^l-\frac13 Q_4^l+\frac{1}{12}Q_6^l-\frac{1}{60}Q_8^l).
\eeq
If the space is bounded at finite proper distance then computing the central charge just amounts to summing the $c^l_{hol}$ contributions from each cell
which represents a non-trivial prediction for the central charge of the 2d CFTs dual to the solutions \eqref{eq:metric}. If one or both of the boundaries are AdS$_4$ then in principle the same is true, albeit now there are an infinite number of cells. We note that for the symmetric profiles discussed earlier we have
\beq
c^{P+n}=c^{P-n-1},~~n=0,...,P-1~~~~\Rightarrow~~~~ \sum_{l=0}^{2P-1}c^l=  2\sum_{l=0}^{P-1}c^l.
\eeq
We will come back to the analysis of the central charge for a concrete choice of solution in section \ref{field-theory}.

\section{A new class of $\mathcal{N}=(0,4)$ $\text{AdS}_3$ solutions to Type IIB}\label{solutionsIIB}

In this section we present a new class of Type IIB solutions consisting on AdS$_3\times$S$^2 \times$S$^2\times$S$^1$ geometries foliated over two intervals. These solutions preserve the $\mathfrak{osp}(4|2)$ superconformal algebra and are generated by acting with Abelian T-duality on the  class of  AdS$_3\times \mathbb{CP}^3$ solutions in \eqref{eq:metric}.\\
~\\ 
The $\mathbb{CP}^3$ space has a natural U(1) isometry on which one can act by T-duality - in  terms of the parametrisation of \eqref{paramCP3} this is $\partial_{\psi}$. 
Following the standard Buscher's procedure leads to a dual metric of the form
\begin{equation}
\begin{split}\label{IIBmetric}
 &\frac{ds^2}{2\pi}=\frac{h}{\sqrt{2hh''-(h')^2}}\,ds^2(\text{AdS}_3)+\frac{2h''}{\sqrt{2hh''-(h')^2}}\,\left[\left(\frac{h'}{h''}-r\right)d\xi-\frac{d\psi}{2\pi\sin\xi\cos\xi}\right]^2\\
 &\qquad+\frac{{\sqrt{2hh''-(h')^2}}}{4h}\, dr^2+\frac{2{\sqrt{2hh''-(h')^2}}}{h''}\left(d\xi^2+\frac14\cos^2\xi ds^2(\text{S}_1^2)+\frac14\sin^2\xi ds^2(\text{S}_2^2)\right)\,. 
 \end{split}
\end{equation}
The dualisation of the rest of the NS-NS sector\footnote{For simplicity of notation we set $l=0$ in \eqref{eq:metric} and \eqref{RRgeometry}.} leads to the following expressions for the dilaton, the $B_2$ field and its field-strength, $H_3=dB_2$,
\begin{equation}
\begin{split}
& e^{\Phi}=\frac{2}{\sin\xi\cos\xi\,h''}\,,\\
&B_2=\pi \left(\frac{h'}{h''}-r \right)\,\bigl[-\cos^2\xi\,\text{vol}(\text{S}_1^2)+\sin^2\xi\,\text{vol}(\text{S}_2^2)\bigr]+\left(\eta_1+\eta_2\right)\wedge d\psi\,,\\
&H_3=2\pi\sin\xi\cos\xi\left[\left(\frac{h'}{h''}-r\right)d\xi-\frac{d\psi}{2\pi\sin\xi\cos\xi}   \right]\wedge\bigl[\text{vol}(\text{S}_1^2)+\text{vol}(\text{S}_2^2)\bigr]\\
&\qquad+\frac{\pi h'h'''}{(h'')^2}\,dr\wedge\bigl[\cos^2\xi\,\text{vol}(\text{S}_1^2)-\sin^2\xi\,\text{vol}(\text{S}_2^2)\bigr]\,.
\end{split}
\end{equation}
Finally, the R-R fluxes are given by
\begin{equation}
\begin{split}\label{RRIIBfluxes}
& F_1=\sin\xi\cos\xi\,\left(h''-r h''' \right)d\xi-\frac{h'''}{2\pi}\,d\psi\,,\\
& F_3=\pi\sin\xi\cos\xi\,F_{3,\xi\psi}\wedge\bigl[\cos^2\xi\,\text{vol}(\text{S}_1^2)-\sin^2\xi\,\text{vol}(\text{S}_2^2)\bigr]\,,\\
&F_5=2\pi^2\sin\xi\cos\xi F^{e}_{5,\xi\psi}\wedge\text{vol}(\text{AdS}_3)\wedge dr+\pi^2\sin^3\xi\cos^3\xi \,F^{m}_{5,\xi\psi}\wedge \text{vol}(\text{S}_1^2)\wedge \text{vol}(\text{S}_2^2)\,,
\end{split}\,
\end{equation}
where we introduced the quantities $F_{3,\xi\psi}$, $F^{e}_{5,\xi\psi}$ and $F^{m}_{5,\xi\psi}$ given by
\begin{equation}
 \begin{split}
 &F_{3,\xi\psi}=\frac{h'(h''+rh''')-r(h'')^2}{h''}\,d\xi-\frac{(h'')^2-h'h'''}{h''}\,\frac{d\psi}{2\pi\sin\xi\cos\xi}\,,\\
 &F^{e}_{5,\xi\psi}=\left(rh'-2h-\frac{hh'(h'-rh'')}{2hh''-(h')^2} \right)'d\xi+\frac12\,\left( 3h'+\frac{(h')^3}{2hh''-(h')^2} \right)'\frac{d\psi}{2\pi\sin\xi\cos\xi}\,,\\
 &F^{m}_{5,\xi\psi}=\frac{-6h(h'')^2+3(h')^2h''+r(h')^2h'''}{(h'')^2}\,d\xi+\frac{(h')^2h'''}{(h'')^2}\,\frac{d\psi}{2\pi\sin\xi\cos\xi}\,.\\
 \end{split}
\end{equation}
The function $h$ satisfies the same master equation as the AdS$_3\times \mathbb{CP}^3$ Type IIA solutions given by \eqref{eq:metric}, namely
\begin{equation}
 h'''=-2\pi F_0\,,
 \end{equation}
where in Type IIB $F_0$ is now associated to the charge of D7 branes, with a discontinuous $F_0$ leading to sources smeared along the $\psi$ direction. 
 
As for its dual partner \eqref{eq:metric}, finding a brane intersection reproducing the background \eqref{IIBmetric} as a near-horizon geometry is particularly challenging. To this end one can try, at least, to differentiate the defect branes, associated to the back-reacted AdS$_3$ geometry, and the ``mother branes", associated to the ambient theory, including the defect.  As we already argued in the previous section the mother theory is ABJM. On the supergravity side, this is realised when one fixes $F_0=0$ in  \eqref{eq:metric}. An analogous AdS$_4$  limit can also be obtained in Type IIB, which is actually the context in which the ABJM brane intersection was first discussed: This consists of a stack of D3 branes stretched between NS5' and $(1, k)5'$ branes, as depicted in Table \ref{tableABJM}. These constitute the mother branes.
In the next section we will propose a concrete brane intersection consisting on NS5-D7 defect branes intersecting the ABJM bound state breaking the isometries of the AdS$_4\times \mathbb{CP}^3$ vacuum. Our concrete brane picture will be obtained by studying the Page charges.
 
 The main elements that make the construction of the exact brane solution associated to our backgrounds particularly intricate is that it is not possible to extract the AdS$_4$ vacuum as a local, asymptotic limit of the AdS$_3\times \text{I}$ solutions. Rather, as explained in the previous section, it arises through a global gluing procedure. This is the manifestation of the fact that the back-reaction of the defect branes irredeemably breaks  the AdS$_4$ vacuum geometry. This implies that we cannot study the equations of motion of defect and mother branes  independently, as it can be done for other simpler AdS$_3$ solutions. Another difficulty is the presence of fractional branes. These are produced by dualising $H_3$ and the magnetic 4-flux in \eqref{eq:metric}, which were defined along 2- and 4-cycles within the $\mathbb{CP}^3$ space. From the Type IIA perspective these fluxes were associated to fractional D4 branes.

In the remainder of this section we identify the sub-class of Type IIB solutions associated to the ABJM theory. Fixing
\begin{equation}
 h'''=0\,~~~~\Rightarrow~~~~  h=c_2+c_4 r+\frac{c_6}{2}\,r^2
\end{equation}
again implies that the $(\text{AdS}_3,r)$ directions become AdS$_4$, as explained in section \ref{ABJM}. This is equivalent to taking the T-dual of the AdS$_4\times \mathbb{CP}^3$ solution. In this limit supersymmetry is enhanced and the background \eqref{IIBmetric} takes the form of an $\mathcal{N}=4$ solution with topology $\mrm{AdS}_4\times \text{S}^2\times \text{S}^2\times \Sigma_2$, which were classified in \cite{DHoker:2007zhm}. 
Specifically, we find the metric
\begin{equation}
 \begin{split}
  \frac{ds^2}{2\pi}&=\frac{\sqrt{2c_2c_6-c_4^2}}{2c_6}\,\left(ds^2(\text{AdS}_4)+4d\xi^2+\cos^2\xi ds^2(\text{S}_1^2)+\sin^2\xi ds^2(\text{S}_2^2)\right)\\
  &\quad+\frac{2c_4^2}{c_6\sqrt{2c_2c_6-c_4^2}}\,\bigl(d\xi-\frac{c_6}{2\pi c_4\sin\xi\cos\xi}\,d\psi\bigr)^2. 
 \end{split}
\end{equation}
From this expression we can extract the 2d metric over the Riemann surface $\Sigma_2$, parametrised by the coordinates $(\xi, \psi)$. We observe that this surface is an annulus.

\noindent The rest of the NS-NS sector boils down to 
\begin{equation}
\begin{split}\label{NSNSABJMlimit}
& e^{\Phi}=\frac{2}{c_6\,\sin\xi\cos\xi}\,,\\
&B_2=\frac{\pi c_4}{c_6} \,\bigl[-\cos^2\xi\,\text{vol}(\text{S}_1^2)+\sin^2\xi\,\text{vol}(\text{S}_2^2)\bigr]+\left(\eta_1+\eta_2\right)\wedge d\psi\,,\\
&H_3=\left(\frac{2\pi c_4}{c_6}\sin\xi\cos\xi\,d\xi-d\psi   \right)\wedge\bigl(\text{vol}(\text{S}_1^2)+\text{vol}(\text{S}_2^2)\bigr)\,,
\end{split}
\end{equation}
while the RR fluxes read
\begin{equation}
\begin{split}\label{RRABJMlimit}
& F_1=c_6\,\sin\xi\cos\xi\, d\xi\,,\\
& F_3=\frac{c_6}{2}\,\left(\frac{2\pi c_4}{c_6}\sin\xi\cos\xi\,d\xi-d\psi   \right)\wedge\bigl[\cos^2\xi\,\text{vol}(\text{S}_1^2)-\sin^2\xi\,\text{vol}(\text{S}_2^2)\bigr]\,,\\
&F_5=-\frac{3\pi}{2}\sqrt{2c_2c_6-c_4^2}\,\text{vol}(\text{AdS}_4)\wedge\left(\frac{2\pi c_4}{c_6}\sin\xi\cos\xi d\xi-d\psi\right)\\
&\qquad+\frac{3\pi^2}{c_6}\sqrt{2c_2c_6-c_4^2}\sin^3\xi\cos^3\xi\,d\xi\wedge \text{vol}(\text{S}_1^2)\wedge \text{vol}(\text{S}_2^2).
\end{split}
\end{equation}

We conclude by pointing out that in order to reproduce the ABJM vacuum in the form discussed in \cite{Aharony:2008ug}, we have to perform the following coordinate transformation within the Riemann surface,
\begin{equation}\label{changepsi}
 \psi \to \psi +\frac{2\pi c_4}{c_6}  \sin \xi \cos \xi \,d \xi\,.
\end{equation}
The effect of this transformation is to turn off the components of $H_3$, $F_3$ and $F_5$ in \eqref{NSNSABJMlimit}  and \eqref{RRABJMlimit} orthogonal to $\psi$. These flux terms are pure gauge and they are typically produced when one T-dualises a solution with an exact $B_2$ field, like our Type IIA solution \eqref{eq:metric}.

\section{The dual field theory}\label{field-theory}

In this section we take the first steps towards identifying the field theory dual to the solutions in the general massive case. As in \cite{Aharony:2008ug,Aharony:2008gk} we propose a brane set-up  in Type IIB string theory. We argue that this brane set-up preserves $\mathcal{N}=(0,3)$ supersymmetry, that should be enhanced to $(0,6)$ in the IR. The mechanism at work should be a 2d realisation of the supersymmetry enhancement that occurs in the ABJM/ABJ theory, albeit with half the supersymmetries. In this case the enhancement is to manifest $(0,4)$ supersymmetry in Type IIB,  consistently with the number of supersymmetries preserved by the Type IIB $\text{AdS}_3$ solutions just discussed. This becomes $(0,6)$ when the T-duality is undone to recover the Type IIA description, again in agreement with the number of supersymmetries preserved by the Type IIA solutions.

\subsection{Hanany-Witten brane set-up}\label{HWset-up}

Let us first propose the brane set-up. On top of the D2, KK and D6 branes (or D3, NS5$'$ and D5$'$-branes in the Type IIB description) of the massless case, depicted in Table \ref{tableABJM}, there are now additional D8 and NS5 branes (or D7 and NS5 branes in the Type IIB description), as well as D4 (or D5) fractional branes. 
A natural interpretation of the
$\text{AdS}_3\times \mathbb{CP}^3$ solutions is then as the backreacted geometries that arise when a large number of D8-NS5 defect branes are introduced in the ABJ theory, in a way that preserves $\mathcal{N}=(0,6)$ supersymmetry, that is, half of the supersymmetries, and a OSp$(6|2)$ superconformal group. The NS5-branes have the effect of bounding the D2 and D6 branes along the domain wall direction of the D8 branes. 


In what follows we will consider the brane intersection in Type IIB, after a T-duality is performed along the Hopf fibre of the S$^3$ contained in the $\mathbb{CP}^3$, in the parametrisation  \eqref{paramCP3}. As discussed in section \ref{solutionsIIB}, the T-duality breaks the supersymmetry from $\mathcal{N}=(0,6)$ to $\mathcal{N}=(0,4)$. In the brane intersection the 8 (Poincar\'e) supersymmetries preserved by the D3-NS5$'$-D5$'$ branes of the ABJM set-up prior to the rotation of the D5$'$-branes are broken by a half by the NS5-branes, giving rise to $(0,4)$ supersymmetry in 2d\footnote{This was shown explicitly in \cite{Chung:2016pgt}.}. The D7-branes are added without breaking any further supersymmetries. As in ABJM, once the rotation of the D5$'$-branes is taken into account supersymmetry should be further broken to $(0,3)$, if the branes are rotated the same angle along the $[3,7]$, $[4,8]$ and $[4,9]$ directions with respect to the NS5$'$-branes\footnote{Otherwise the supersymmetries would be reduced to $(0,2)$.}.

The Type IIB brane configuration is depicted in Table \ref{HWbranesetupIIB}.
\begin{table}[http!]
\renewcommand{\arraystretch}{1}
\begin{center}
\scalebox{1}[1]{
\begin{tabular}{c| c cc  c c  c  c c c c}
 branes & $x^0$ & $x^1$ & $r$ & $x^3$ & $x^4$ & $x^5$ & $\psi$ & $x^7$ & $x^8$ & $x^9$ \\
\hline \hline
$\mrm{D}3$ & $\times$ & $\times$ & $\times$ & $-$ & $-$ & $-$ & $\times$ & $-$ & $-$ & $-$ \\
$\mrm{NS}5'$ & $\times$ & $\times$ & $\times$ & $\times$ & $\times$ & $\times$ & $-$ & $-$ & $-$ & $-$ \\
$(1,k) 5'$ & $\times$ & $\times$ & $\times$ & $\cos{\theta}$ & $\cos{\theta}$ & $\cos{\theta}$ & $-$ & $\sin{\theta}$ & $\sin{\theta}$ & $\sin{\theta}$ \\
$\mrm{D}7$ & $\times$ & $\times$ & $-$ & $\times$ & $\times$ & $\times$ & $-$ & $\times$ & $\times$ & $\times$ \\
$\mrm{NS}5$ & $\times$ & $\times$ & $-$ & $-$ & $-$ & $-$ & $\times$ & $\times$ & $\times$ & $\times$ \\
\end{tabular}
}
\caption{Brane intersection describing the ABJ brane system in Type IIB with D7-NS5 ``edge'' states. The intersection describes a brane box with D3 colour branes extended along the $r$ and $\psi$ directions. The brane intersection should preserve $\mathcal{N}=(0,3)$ supersymmetry in 2d.} \label{HWbranesetupIIB}
\end{center}
\end{table}
In this brane set-up the D7-NS5 defect branes create a domain wall in the 3d theory living in the D3-NS5$'$-$(1,k) 5'$ branes. This is similar to the D4-D8 defect branes  introduced  in \cite{Fujita:2009kw} as probe defect branes in ABJM,  in order to realise edge states in the Fractional Quantum Hall Effect.  
The effect of these branes in the probe brane approximation was to modify the ranks and levels of the gauge groups of the Chern-Simons theory living in the D3-branes. Instead, our study shows that for a large number of NS5-D8 defect branes the resulting field theory becomes two dimensional, while still preserving one half of the supersymmetries and a subgroup of the superconformal group. 

\begin{figure}[h]
\centering
\includegraphics[scale=0.6]{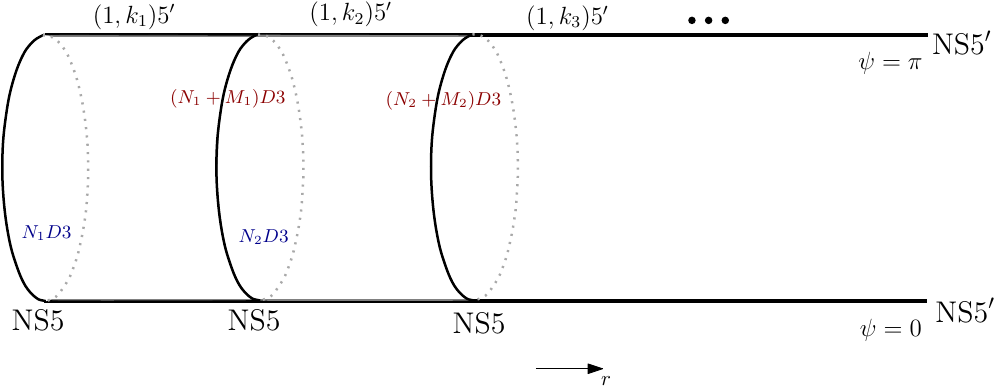}
\caption{Brane configuration associated to the $\text{AdS}_3\times\mathbb{CP}^3$ solutions. $N_l$ D3-branes stretch between $\psi=0$ and $\pi$ and $N_l+M_l$ between $\psi=\pi$ and $2\pi$.}\label{branediagram}
\end{figure}

\begin{figure}[h]
\centering
\includegraphics[scale=0.6]{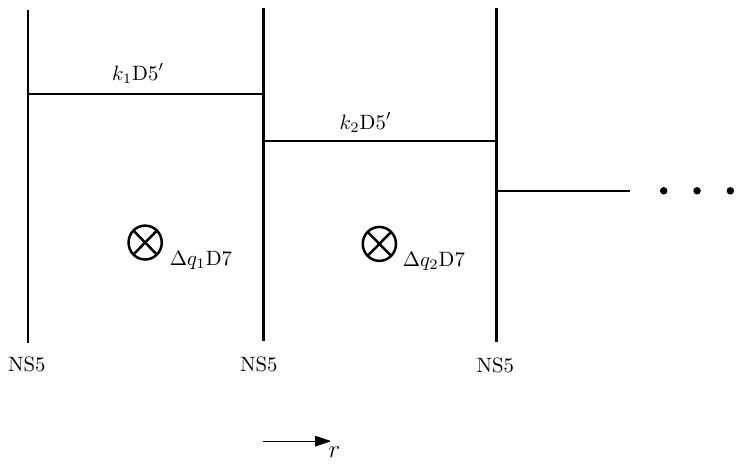}
\caption{Brane creation effect for the D5$'$-NS5-D7 subsystem of the brane set-up.}\label{D5-D7}
\end{figure}

In more detail, the brane set-up depicted in Table \ref{HWbranesetupIIB} describes a brane box model in which $N_l$ D3-branes are bounded between NS5-branes along the $r$ direction, and between the NS5$'$ brane located at $\psi=0$ and the $(1,k_l)$ 5$'$ branes (a bound state of 1 NS5$'$ and $k_l$ D5$'$ branes) located at $\psi=\pi$ along the $\psi$-direction. This number becomes $N_l+M_l$ between the $(1,k_l)$ 5$'$-branes and the NS5$'$-brane at $\psi=2\pi$, due to extra fractional D5-branes created between these branes. We have depicted the brane picture in Figure \ref{branediagram}.
In this configuration the number of D5$'$-branes  bounded with the NS5$'$ brane at $\psi=\pi$, $k_l$, changes as one moves along the $r$-direction and therefore also the rotation angle $\theta_l$ between both types of branes\footnote{Even if we have not indicated this explicitly in the drawing for the sake of clarity.}, whose tangent is given by $\tan{\theta_l}=\Delta k_l$. This happens due to a brane creation effect when the D7 flavour branes are crossed. We have depicted this part of the brane set-up  in Figure \ref{D5-D7}. This subsystem 
describes a D5$'$-D7-NS5 brane intersection with five common worldvolume directions, $k_l$ D5$'$-branes stretched between NS5-branes and $\Delta q_l$ D7 flavour branes located at each $r\in [l,l+1]$ interval. As we will see in the next subsection the number of D5$'$-branes created across intervals is the one that corresponds to balanced 5d quivers, as described in \cite{Legramandi:2021uds}.

The whole configuration is thus a brane-box model in which D3 colour branes are stretched between NS5-branes along the $r$-direction and between NS5$'$ and $(1,k_l)$ 5'-branes along the $\psi$-direction, with the number of D5$'$-branes changing across intervals due to the presence of the D7-branes. On top of this, D5 fractional branes are created in the $\psi\in [\pi,2\pi]$ interval when the $k_l$ D5$'$-branes are crossed. 

In the next subsection we analyse the field content associated to this configuration and propose a concrete quiver field theory that describes it.

\subsection{Building blocks}

As in 3d, the $\mathcal{N}=(0,3)$ gauge theory living in the D3-branes is expected to have the same field content as that of an $\mathcal{N}=(0,4)$ gauge theory, except for the deformations introduced by the rotations of the 5-branes, responsible for the breaking of the supersymmetry to  $\mathcal{N}=(0,3)$. Specifically, the $(0,4)$ theory has SO(4) R-symmetry, that should reduce to SU(2)$_R=\text{diag} (\text{SU}(2)\times \text{SU}(2))$ for the $(0,3)$ theory. In 3d the $\mathcal{N}=4$ Coulomb branch combines with the $\mathcal{N}=4$ Higgs branch such that $\mathcal{N}=3$ supersymmetry remains. In 2d $(0,4)$ theories there is no Coulomb branch, because the scalars live now in (twisted) adjoint hypermultiplets. The mixing between the Coulomb and Higgs branches should then be replaced by a combination between the twisted and untwisted $(0,4)$ hypermultiplets, resulting in $(0,3)$ supersymmetry. Unfortunately, since not much is known about 2d $(0,3)$ theories, in this subsection we will take a ``phenomenological'' approach, hoping that our discussion will stimulate more detailed investigations.

We start recalling the field content of the $(0,4)$ brane box models studied in \cite{Hanany:2018hlz}, to which we will add the rotation of the D5$'$-branes relative to the NS5$'$-branes.


The brane box model studied in \cite{Hanany:2018hlz} consists on a D3-NS5-NS5$'$-D5-D5$'$ brane intersection in which the branes are oriented as depicted in Table \ref{HO}, with the whole brane system preserving large $\mathcal{N}=(0,4)$ supersymmetry. 
\begin{table}[http!]
\renewcommand{\arraystretch}{1}
\begin{center}
\scalebox{1}[1]{
\begin{tabular}{c| c cc  c c  c  c c c c}
 branes & $x^0$ & $x^1$ & $r$ & $x^3$ & $x^4$ & $x^5$ & $\psi$ & $x^7$ & $x^8$ & $x^9$ \\
\hline \hline
$\mrm{D}3$ & $\times$ & $\times$ & $\times$ & $-$ & $-$ & $-$ & $\times$ & $-$ & $-$ & $-$ \\
$\mrm{NS}5'$ & $\times$ & $\times$ & $\times$ & $\times$ & $\times$ & $\times$ & $-$ & $-$ & $-$ & $-$ \\
$\mrm{D}5'$ & $\times$ & $\times$ & $\times$ & $-$ & $-$ & $-$ & $-$ & $\times$ & $\times$ & $\times$ \\
$\mrm{NS}5$ & $\times$ & $\times$ & $-$ & $-$ & $-$ & $-$ & $\times$ & $\times$ & $\times$ & $\times$ \\
$\mrm{D}5$ & $\times$ & $\times$ & $-$ & $\times$ & $\times$ & $\times$ & $\times$ & $-$ & $-$ & $-$ \\
\end{tabular}
}
\caption{Brane intersection describing the D3-brane box model studied in \cite{Hanany:2018hlz}, where we have denoted $r$ and $\psi$ the field theory directions, as in the brane set-up depicted in Table \ref{HWbranesetupIIB}.} \label{HO}
\end{center}
\end{table} 
In this set-up the SO(3)$_{345}\times$SO(3)$_{789}$ rotation group is identified with the SO(4)$_R$ R-symmetry. In our case the NS5$'$ and the D5$'$ branes are located at the same position in $\psi$ and are rotated the same angle on the $[3,7]$, $[4,8]$ and $[5,9]$ directions. These rotations render some of the scalars of the brane box massive \cite{Kitao:1998mf,Bergman:1999na}, as we discuss below.

\begin{figure}[h]
\centering
\includegraphics[scale=0.75]{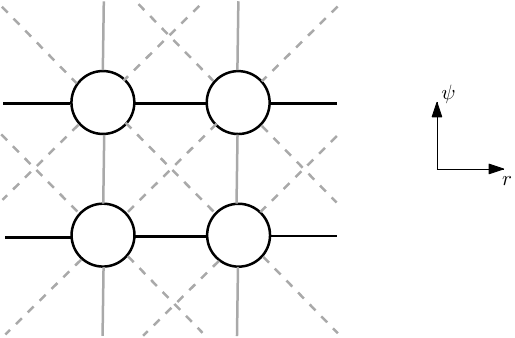}
\caption{General quiver associated to D3-branes stretched between NS5 and NS5' branes in two perpendicular directions. Circles denote $(0,4)$ vector multiplets, black lines $(0,4)$ twisted hypermultiplets, grey lines $(0,4)$ hypermultiplets and grey dashed lines $(0,2)$ Fermi multiplets.}\label{HOquiver}
\end{figure}

Prior to the rotation of the D5$'$-branes the quantisation of the open strings with ends on the D3-branes in the brane box model gives rise to four types of $\mathcal{N}=(0,4)$ multiplets \cite{Hanany:2018hlz}:
\begin{itemize}
\item When the end-points of the string lie on the same stack of D3-branes the projections induced by both the NS5 and NS5$'$ branes leave behind a $(0,4)$ vector multiplet.
\item When the end-points of the string lie on two different stacks of D3-branes separated by an NS5-brane the degrees of freedom along the $(x^7,x^8,x^9)$ directions are fixed, leaving behind the scalars associated to the $(x^3,x^4,x^5)$ directions, which together with the 
$A_r$ component of the gauge field give rise to a $(0,4)$ twisted hypermultiplet in the bifundamental representation. The hypermultiplets are twisted because the scalars are charged with respect to the SO(3)$_{345}$ subgroup of the R-symmetry group.
\item When the end-points of the string lie on two different stacks of D3-branes separated by an NS5$'$-brane the degrees of freedom along the $(x^3,x^4,x^5)$ directions are fixed, leaving behind the scalars associated to the $(x^7,x^8,x^9)$ directions, which together with the 
$A_\psi$ component of the gauge field give rise to a $(0,4)$ hypermultiplet in the bifundamental representation. The hypermultiplets are untwisted because the scalars are charged with respect to the SO(3)$_{789}$ subgroup of the R-symmetry group.
\item When the end-points of the string lie on two different stacks of D3-branes separated by both an NS5 and an NS5$'$ brane all the scalars are fixed, leaving behind the fermionic mode associated to a bifundamental $(0,2)$ Fermi multiplet.
\end{itemize}
We have depicted the building blocks of the $(0,4)$ quivers just described in Figure \ref{HOquiver}. In this quiver circles denote $(0,4)$ vector multiplets, black lines $(0,4)$ twisted hypermultiplets, grey lines $(0,4)$ hypermultiplets and grey dashed lines $(0,2)$ Fermi multiplets.
To the previous matter content  we have to add the  multiplets that come from the D5$'$ flavour branes\footnote{Since the D5-branes are fractional branes.}. These branes contribute with $(0,4)$ fundamental hypermultiplets to the D3-branes lying in the same $r$-interval, and with $(0,2)$ fundamental Fermi multiplets to the D3-branes in adjacent $r$-intervals. On top of this we need to add the contribution of the D7-branes, which being completely orthogonal to the D3-branes just contribute with $(0,2)$ fundamental Fermi multiplets to the D3-branes lying in the same $r$-interval.


\subsection{2d quivers}

\begin{figure}[h]
\centering
\includegraphics[scale=0.55]{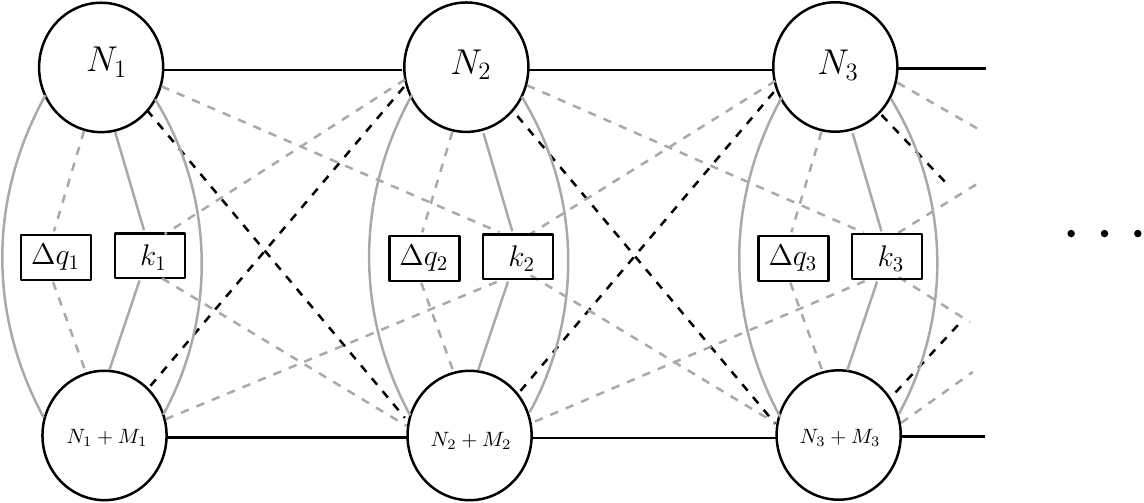}
\caption{Quiver diagram associated to our brane set-up. Black dashed lines denote $(0,4)$ Fermi multiplets.}\label{HOenhanced}
\end{figure}

\begin{figure}[h]
\centering
\includegraphics[scale=0.55]{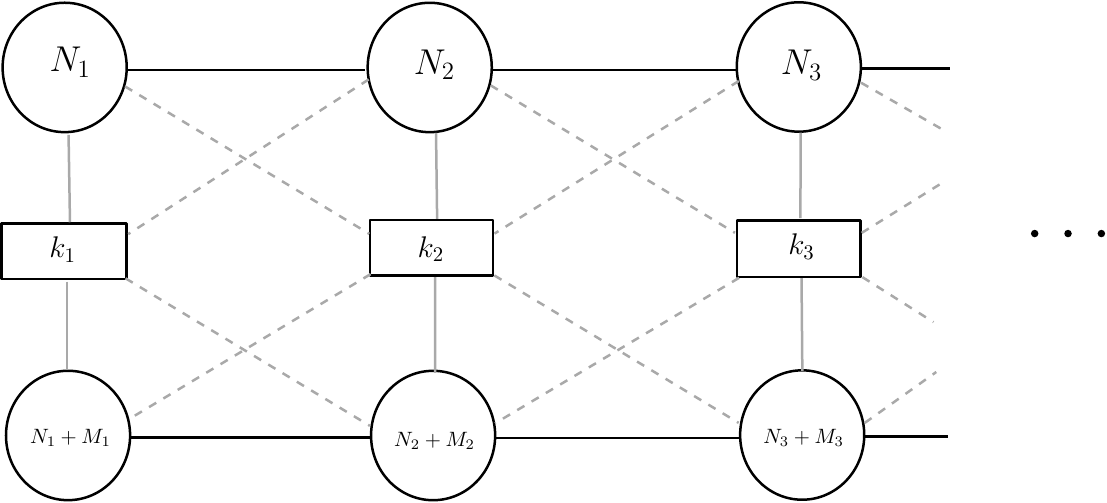}
\caption{Quiver diagram showing the matter fields coming from the open strings stretched between D3-branes in the same $\psi$-interval and D3-branes and D5'-branes.}\label{subquiver1}
\end{figure}

\begin{figure}[h]
\centering
\includegraphics[scale=0.55]{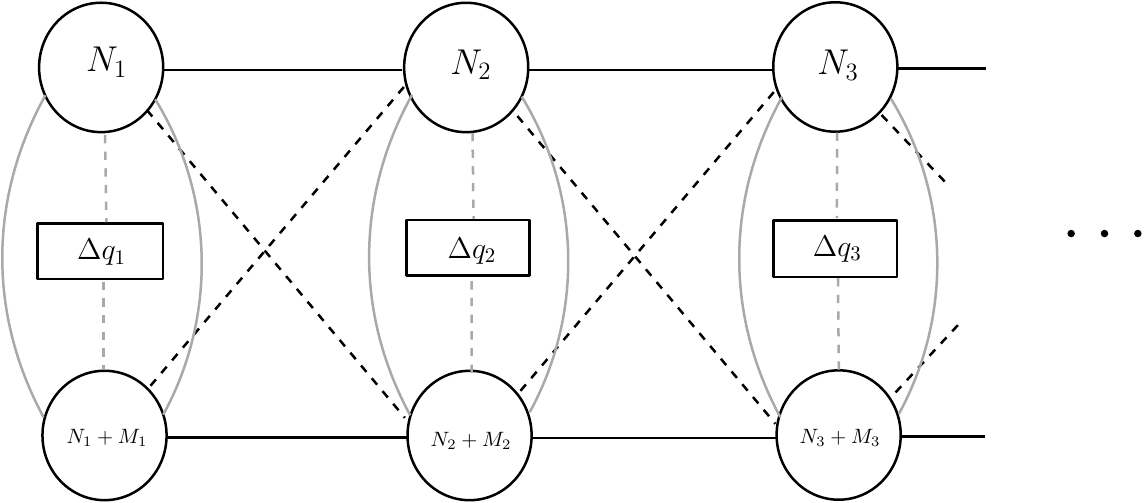}
\caption{Quiver diagram showing the matter fields coming from the open strings stretched between D3-branes in the same and different $\psi$-intervals and D3-branes and D7-branes.}\label{subquiver2}
\end{figure}

Taking into account the field content arising from the different branes in the brane box configuration just reviewed we can now build up the quiver associated to our particular brane set-up.  We have depicted it in Figure \ref{HOenhanced}. 
For the sake of clarity we have decomposed the matter content of this quiver in the two quivers depicted in Figures \ref{subquiver1} and \ref{subquiver2}.
Since now $\psi$ is a compact direction, there are two $(0,4)$ hypermultiplets connecting the $N_l$ and $N_l+M_l$ gauge groups at each $r$-interval, that originate from the open strings that connect the $N_l$ D3 and $N_l+M_l$ D3 branes across either one of the NS5$'$ branes. Similarly, there are two $(0,2)$ Fermi multiplets connecting $N_l$ D3-branes with $N_{l'}+M_{l'}$ D3 branes in adjacent $r$-intervals, depending on the NS5$'$-brane crossed by the open strings. We have denoted these $(0,4)$ Fermi-multiplets with black dashed lines. In each $r$-interval the two $(0,4)$ bifundamental hypermultiplets connecting the gauge nodes with ranks $N_l$ and $N_l+M_l$  combine onto a $(0,6)$ bifundamental hypermultiplet. On top of this we have the $(0,4)$ fundamental hypermultiplets and $(0,2)$ fundamental Fermi multiplets associated to the open strings that connect the D3-branes with the D5$'$-branes and the D7-branes. In all, our quiver field theory consists on a sequence of U$(N_l)$ and U$(N_l+M_l)$ gauge groups with the field content of $(0,4)$ vector multiplets connected to each other by $(0,6)$ bifundamental hypermultiplets, with extra $k_l$ fundamental or antifundamental $(0,4)$ hypermultiplets\footnote{Taking into account that due to the different relative positions of the NS5$'$-branes at $\psi=0,2\pi$ compared to the $(1,k_l)$ 5-branes at $\psi=\pi$, the flavour groups contribute with fundamentals to the $N_l$ gauge nodes and with antifundamentals to the $N_l+M_l$ gauge nodes.}. Note that at this point we have not yet taken into account the rotation between the D5$'$ and the NS5$'$-branes. We will do this after we discuss gauge anomaly cancellation.

Taking into account the contribution of the different multiplets to the gauge anomaly, summarised in Table \ref{Table:gaugeanom}, 
\begin{table}[h]
\begin{center}
\begin{tabular}{|c|c|}
\hline
Multiplet & Contribution\\
\hline\hline
(0,4) hyper & 1 \\
\hline
(0,4) vector  & -2N \\
\hline
(0,2) Fermi& $-\frac12$ \\
\hline
\end{tabular}
\end{center}
\caption{Contribution to the gauge anomaly of the different multiplets that couple in the quiver depicted in Figure \ref{HOenhanced}.}
\label{Table:gaugeanom}
\end{table}
we find that the following condition must be satisfied for each node in the quiver 
\begin{equation}
2M_l-M_{l-1}-M_{l+1}+k_l-\frac12 k_{l-1}-\frac12 k_{l+1}-\frac12 \Delta q_l=0. \label{anomaly}
\end{equation}

We proceed now with checking whether this is satisfied by the concrete quivers associated to our solutions.

In doing this we first need to recall that due to the effects of the Freed-Witten anomaly and the higher curvature terms  the numbers of branes in the brane set-up are not the same as the quantised charges built from the solutions, as discussed already for the massless case in section \ref{ABJM}. In the massive case the introduction of D8(D7)-branes in the different intervals produces further changes to the ones already summarised in the massless case, given by \eqref{change}.
As shown in  \cite{Bergman:2010xd}, the introduction of D8(D7)-branes produces a jump on the field theory parameters due to the brane creation effect, but in non-trivial topologies introduces as well extra D4(D5)-brane charge coming from the half-integer component of the $B_2$ flux needed to cancel the Freed-Witten anomaly and the higher curvature contributions. Through an analogous calculation to the one that shows that a D6-brane wrapped on $\mathbb{CP}^3$ carries 1/12 units of D2-brane charge, as a result of the sum of a contribution from a Chern-Simons term and a higher curvature term, one finds that a D8-brane domain wall carries 1/12 units of D4-brane charge dissolved in its worldvolume. Putting all this together one finds that the
field theory D2, D4, D6 and D8 brane charges in the different $r$-intervals, denoted by $(N_l, M_l, k_l, q_l)$, are related to the $(Q_2^l, Q_4^l, Q_6^l, Q_8^l)$ quantised charges computed in equations \eqref{Qs} through
\begin{eqnarray}
&&Q_2= N+\frac{k}{12}, \nonumber\\
&&Q_4=M-\frac{k}{2}+\frac{q}{12}, \nonumber\\
&&Q_6= k, \nonumber\\
&&Q_8=-q, \label{shifted}
\end{eqnarray}
where we have used Type IIA notation to make the comparison with the corresponding expressions in the massless case easier. These relations were already found in  \cite{Bergman:2010xd}  in a non-supersymmetric massive extension of the ABJ theory considered therein. 
Using these relations one can then see that the transformations \eqref{changeQs} give rise to the following transformations for the numbers of D3 (D2), D5 (D4) and D5$'$ (D6) branes across intervals
\begin{eqnarray}
&&N_l= N_{l-1}-M_{l-1}+k_{l-1}, \label{Seibergm2}\\
&&M_l+\frac{q_l}{12}=M_{l-1}+\frac{q_{l-1}}{12}-k_{l-1},\label{Seibergm3} \\
&&k_l=k_{l-1}+q_{l-1},\label{Seibergm}
\end{eqnarray}
where we recall that even if in our configuration the D5-branes are fractional branes, their numbers are still changing across intervals. 
Consistently with this, one can check that the condition \eqref{Seibergm3}  is the same needed to ensure that an NS5-brane is created across intervals, following the same reasoning around equation \eqref{NS5creation}, now in terms of the numbers of D4(D5), NS5 and D8(D7)-branes. 
Note that the change in the D4(D5)-brane charge is due, on the one hand, to the jump produced by the D8(D7)-brane domain wall, and, on the other, to the D4(D5)-brane charge dissolved in its worldvolume, due to the topological CS and higher curvature terms. The change in the D4(D5)-brane charge due to the brane creation effect is the shift by $-k_{l-1}$.

As we already mentioned, we can understand the change of the D5$'$-brane charge across intervals by looking at the 5d field theory living in the D5$'$-D7-NS5 brane subsystem of the brane set-up. The D7-branes introduced between consecutive NS5-branes at $r=l, l+1$ create $k_l$ D5$'$-branes stretched between the NS5-branes, satisfying the condition of balanced 5d quivers \cite{Legramandi:2021uds}
\begin{equation}\label{anomaly2}
2 k_l=k_{l-1}+k_{l+1}+\Delta q_l,
\end{equation}
with $\Delta q_l=q_{l-1}-q_l$ the number of D7-branes in the $[l,l+1]$ interval, responsible for the jump in the charge from $q_{l-1}$ to $q_l$. This is precisely as implied by equation \eqref{Seibergm}. As we mentioned, the D5$'$-branes become however flavour in the complete brane intersection, being orthogonal to the D3-branes along the $\psi$-direction. 

In turn, the D5-brane charge generated by brane creation, that is, excluding the charge dissolved in the D7-branes, satisfies
\begin{equation}\label{anomaly3}
2M_l=M_{l-1}+M_{l+1}-\Delta k_l,
\end{equation}
with $\Delta k_l=k_{l-1}-k_l$. This condition shows that the number of D5 fractional branes changes across intervals due to the D5$'$-branes that are being created at each $r$ interval because of the D7-branes.

At this point we can now proceed with checking equation \eqref{anomaly}. 
Substituting \eqref{anomaly2} this condition becomes
\begin{equation}
2M_l-M_{l-1}+M_{l+1}=0,
\end{equation}
which clearly is only satisfied by our quantised charges when $\Delta k_l=0$, due to equation \eqref{anomaly3}.
Therefore, as expected, we find that only in the massless case, when the quivers are truly $(0,4)$ supersymmetric\footnote{They are actually $\mathcal{N}=3$ in 3d, as discussed in section \ref{ABJM}.}, the condition for gauge anomaly cancellation for $(0,4)$ quivers is satisfied. Instead, for our solutions the D5$'$-branes are rotated with respect to the NS5$'$-branes, and this renders the scalar fields transverse to the D5$'$-branes massive \cite{Bergman:1999na,Kitao:1998mf}. The mass of these fields is proportional to the tangent of the rotation angle between the branes, given at each interval by $\tan{\theta_l}=\Delta k_l$, which is precisely the extra term in equation \eqref{anomaly3}. This shows that in the $(0,3)$ theory, when the massive scalar fields are not taken into account in the calculation, gauge anomaly cancellation is indeed satisfied. This provides a non-trivial consistency check for our proposed quivers and their preserved supersymmetries.

Having checked the consistency of our proposal we construct in the next subsection a concrete completion of the quivers depicted in Figure \ref{HOenhanced}.

\subsubsection{Completion of the quivers}

\begin{figure}[h]
\centering
\includegraphics[scale=0.9]{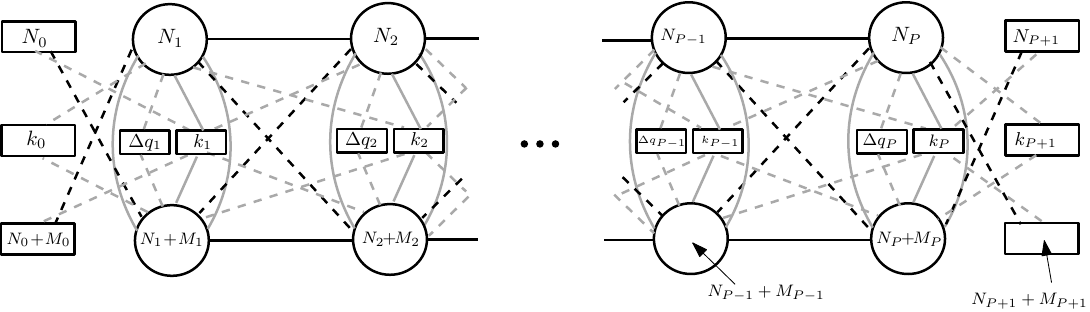}
\caption{Quiver completed by two AdS$_4$ regions for $r<1$ and and $r\geq P$ .}\label{completedquiver}
\end{figure}

In order for our proposed quivers to describe well-defined 2d field theories we have to properly account for how they start and terminate along the $r$ non-compact direction. We will do this by embedding them onto the 3d ABJ theory, that is, by ending the space with two massless regions that are locally $\text{AdS}_4\times \mathbb{CP}^3$, as described in subsection \ref{ABJM}.  As discussed in the previous section, the simplest way to do this geometrically is to only have non trivial $Q^8_l<0$   for $1\leq r< P$ assuming continuity conditions of the type \eqref{sumQP} throughout. The result is depicted in Figure \ref{completedquiver}. 
This shows a quiver like the one depicted in Figure \ref{HOenhanced}  terminating with a massless regions without D8-branes. 

In the massless regions at both ends $r$ becomes part of the external space and a 3d theory is deconstructed, their behaviour is analogous so we will only describe the first region explicitly: Here $k_0$ turns on a Chern-Simons level for the 3d gauge groups with ranks $N_0$ and $N_0\pm M_0$, with opposite signs due to the relative positions of the NS5$'$ at $\psi=0,2\pi$ compared to the $(1,k_0)$ 5-branes at $\psi=\pi$. 
Additionally the two $(0,4)$ Fermi multiplets that were connecting gauge nodes across intervals combine with the $(0,6)$ hypermultiplet connecting the $N_0$ and $N_0\pm M_0$ gauge nodes, to form a $(6,6)$ hypermultiplet. In turn, the $(0,4)$ twisted hypermultiplets connecting the $N_0$ and $N_1$ nodes and the $N_0\pm M_0$ and $N_1\pm M_1$ nodes turn onto adjoint twisted hypermultiplets for each of the $N_0$ and $N_0\pm M_0$ gauge nodes. They then combine with the respective $(0,4)$ vector multiplets to form $(4,4)$ vector multiplets. This is completed by the two gauge nodes being connected wit the $k_0$ flavour group with $(4,4)$ fundamental multiplets. Therefore, we find that in the massless regions supersymmetry is enhanced to 8 Poincar\'e supercharges. The rotations between the D5$'$ and NS5$'$-branes, not taken into account in this analysis, finally render the  number of  supersymmetries equal to 6 Poincar\'e supercharges, equivalent to $\mathcal{N}=3$ in 3d, as in the ABJM/ABJ theory.

We show in the next subsection that the mappings across intervals given by equations \eqref{Seibergm2}-\eqref{Seibergm}  find  an interesting interpretation in relation with Seiberg duality in 3d Chern-Simons matter theories. 

\subsection{Connection with Seiberg duality}

The transformations of the field theory charges given by \eqref{Seibergm2}-\eqref{Seibergm}
\begin{equation}
N\rightarrow N-M+k, \qquad M\rightarrow M-k,\qquad k\rightarrow k+q
\end{equation}
due to brane creation across intervals (which in the case of the D4-branes requires subtracting the charge dissolved in the D8-branes) are precisely the extension of the Seiberg dualities
in ABJM/ABJ discussed in \cite{Aharony:2008gk,Aharony:2009fc} to the non-supersymmetric  U$(N+M)_{k}\times$U$(N)_{-k+q}$ Chern-Simons matter theories proposed in \cite{Bergman:2010xd}. The particular set up in \cite{Bergman:2010xd} is a non-supersymmetric one obtained by a D8-brane deformation of the ABJM/ABJ theory, in which the D8-brane is a non-supersymmetric domain wall embedded into $\text{AdS}_4\times \mathbb{CP}^3$  spanning $\text{AdS}_4$ and a five dimensional subspace of $\mathbb{CP}^3$. Treating the D8-branes as probes it was shown that the deformation leaves the metric unchanged and  interpolates between the ABJM/ABJ
background and the non-supersymmetric AdS$_4\times \mathbb{CP}^3$  background of massive  IIA found in \cite{Gaiotto:2009mv}. A generalisation of the brane creation effects discussed in \cite{Aharony:2009fc} to include the D8-branes led them to propose precisely the transformations above as the generalisation of Seiberg duality to the massive case. These transformations map non-supersymmetric Chern-Simons matter theories with gauge groups U$(N+M)_{k}\times$U$(N)_{-k+q}$ and U$(N)_{k+q}\times $U$(N-M+k)_{-k}$  onto each other\footnote{Still, in the absence of supersymmetry it is not possible to conclude that these theories should be  equivalent in the IR.}.

Our results suggest  that when the D8-branes are embedded in the 3d theory in a supersymmetric way, which implies adding as well NS5 branes and fractional D4-branes, one of the 3d field theory directions (the $r$ direction in the parametrisation \eqref{sinh}-\eqref{param}) turns into an energy scale, and generates a flow towards a 2d CFT. Geometrically the backreaction of the D4-D8-NS5 defect branes gives rise to a $(0,6)$ supersymmetric $\text{AdS}_3\times \mathbb{CP}^3$ geometry where $r$ becomes part of the internal space. In this geometric setting it is now possible to perform large gauge transformations along the $r$ direction that create new branes, that precisely realise the extension of Seiberg duality to the massive case proposed in  \cite{Bergman:2010xd}. 

\subsection{A comment on the central charge}

The holographic central charge of the 2d $(0,6)$ SCFTs dual to the solutions with $h$-functions given by \eqref{hprofile} can be obtained substituting in the general expression  \eqref{centralcharge}. For the completed quivers depicted in Figure \ref{completedquiver} we obtain 
\begin{equation}\label{centralchargemassive}
c_{hol}= \frac{1}{2}\sum_{l=0}^P\Bigl(2N_lk_l-M_l^2+M_lk_l-\frac{1}{12}k_l^2+q_l(N_l-\frac12 M_l+\frac{5}{12}k_l-\frac{13}{720}q_l)\Bigr).
\end{equation}
This constitutes a very non-trivial prediction on the field theory side. One can check that in the massless limit \eqref{centralchargemassive} reduces to equation \eqref{centralcharge2d}, later ``reconverted'' onto a 3d expression in \eqref{cholmassless} (and shown to agree with the field theory results using localization \cite{Drukker:2010nc,Herzog:2010hf,Fuji:2011km,Marino:2011eh}). Note that since we have used the shifted values of the charges this expression includes higher derivative corrections. This implies that the predictions on the field theory side should go beyond the planar approximation. 

On the field theory side the central charge can be computed as the central extension of the  $\mathfrak{osp}(6|2)$ superconformal algebra,  given by  \cite{Bershadsky:1986ms}
\begin{equation}\label{ccalgebra}
c_R=\frac{k(3k+13)}{k+3},
\end{equation}
where $k$ is the level of the algebra.
For superconformal algebras with $(0,2)$ supersymmetry, the level of the algebra can be computed 
from the U$(1)_R$ R-symmetry anomaly, using that (see for instance \cite{Tong:2014yna})
\begin{equation}\label{level}
k=\text{Tr}[\gamma_3Q_R^2],
\end{equation}
where $Q_R$ is the R-charge under the U$(1)_R$ R-symmetry group, and the trace is over all Weyl fermions in the theory. Given that the R-symmetry anomaly is a 't Hooft anomaly, the calculation can be performed in the weakly coupled UV description of the field theory. We have argued that in our case we have $(0,3)$ supersymmetry, so this expression, in principle, does not apply. Still, we can check that it is possible to recover the leading terms in equation \eqref{centralchargemassive} when $q=0$, that is, when the field theory is $(0,4)$ supersymmetric, and therefore expression \eqref{level} can be used. Indeed, recalling the R-charges of the fermions in the different multiplets, summarised in Table \ref{Table:R-charges},
\begin{table}[h]
\begin{center}
\begin{tabular}{|c|c|c|}
\hline
Multiplet & chirality & R-charge\\
\hline\hline
(0,4) hyper &  R.H. & -1 \\
\hline
(0,4) twisted hyper & R.H. & 0 \\
\hline
(0,4) vector  & L.H. & 1 \\
\hline
(0,2) Fermi& L.H. & 0 \\
\hline
\end{tabular}
\end{center}
\caption{R-charges and chiralities of fermions in $(0,4)$ multiplets.}
\label{Table:R-charges}
\end{table}
we find
\begin{equation}
k=\sum_{l=0}^P \Bigl(2N_lk_l+M_lk_l-M_l^2\Bigr),
\end{equation}
which matches the first three terms in \eqref{centralchargemassive} upon suitable normalisation. In turn, the terms in \eqref{centralchargemassive} proportional to the mass should originate field theoretically from genuine $(0,3)$ multiplets, whose contribution to the central charge has not, to our knowledge, been worked out in the literature. This remains as an interesting open problem that deserves further investigation. On more general grounds, one could expect that a mixing between the SU(2)$_R$ R-symmetry and the SU(2) global symmetry occurs in the determination of the infrared R-current, which should then be computed via c-extremisation \cite{Benini:2012cz,Benini:2013cda}. Expression \eqref{centralchargemassive} constitutes in this sense a very useful result that can be used as a lead for the field theory investigations.

Finally, in the remainder of this subsection we show that the holographic central charge can be obtained as a product of electric and magnetic charges associated to the solutions. This generalises the results in \cite{Lozano:2020txg,Lozano:2020sae,Lozano:2021rmk} for $\text{AdS}_2$ backgrounds with $\mathcal{N}=4$ supersymmetries to $\text{AdS}_3$, with different number of supersymmetries. This suggests that the findings in \cite{Lozano:2020txg,Lozano:2020sae,Lozano:2021rmk} should hold more broadly.

\subsubsection{The central charge as a product of electric and magnetic charges}

Following \cite{Lozano:2020txg,Lozano:2020sae,Lozano:2021rmk} we define a density of electric and magnetic charges from the electric and magnetic components of the Page fluxes of a given background, as
\begin{eqnarray}
  \mathcal{Q}_{p}^e=\frac{\hat{f}^e_{p+2}}{(2\pi)^{p}}, \qquad  \mathcal{Q}_{p}^m=\frac{\hat{f}^m_{8-p}}{(2\pi)^{7-p}},
\end{eqnarray}
where, as in these references, the electric density will require a regularisation, as it involves an integration over the $\text{AdS}_3$ subspace.

The product of electric and magnetic densities computed as 
\begin{equation}
\label{eq:page/maxwellCC}
     \mathcal{Q}=\int \sum_{k=0}^{4}  (-1)^{k+1}\mathcal{Q}_{(2k)}^e\mathcal{Q}_{(2k)}^m =\frac{1}{(2\pi)^7}\int \sum_{k=1}^{4}  (-1)^{k+1}\hat{f}^e_{2k+2}\hat{f}^m_{8-2k},
\end{equation}     
gives,
\begin{equation}
\mathcal{Q}=\int \frac{\text{vol}(\text{AdS}_3)}{6\pi}\wedge \frac{\text{vol}(\mathbb{CP}^3)}{\pi^3/6}\wedge dr \left(2hh''-(h')^2+\partial_r\left(hh'-\frac{h'''h^3}{2hh''-(h')^2}\right)\right),
\end{equation} 
upon substitution of the magnetic Page fluxes given by \eqref{eq:pagefluxes} and the corresponding electric components, given by\footnote{For simplicity of notation we set $r=l$.} 
\begin{gather}
\begin{align}
	&\hat{f}^e_4=\frac{- h\pi}{2h h''\!-\!(h')^2} \left(\frac{(h')^3h'''}{2h h''\!-\!(h')^2}\!-\!3(h'')^2\right)\text{vol}(\text{AdS}_3)\!\wedge \! dr,\\[2mm]
	&\hat{f}^e_6=\frac{4 h\pi^2}{2h h''\!-\!(h')^2}\left(\frac{2h(h')^2h'''}{2h h''\!-\!(h')^2}\!-\!3h'h''\right)\text{vol}(\text{AdS}_3)\!\wedge \! J\!\wedge \! dr,\nonumber\\
	&\hat{f}^e_8=\frac{-8 h\pi^3}{2h h''\!-\!(h')^2}\left(\frac{4h'h'''h^2}{2h h''\!-\!(h')^2}\!-\!2(h h''\!+\!(h')^2)\right)\text{vol}(\text{AdS}_3)\!\wedge \!J\!\wedge \! J\!\wedge \! dr,\nonumber\\
	&\hat{f}^e_{10}=\frac{32 h\pi^4}{2h h''\!-\!(h')^2}\left(\frac{8h^3h'''}{3(2h h''\!-\!(h')^2)}\!-\!2hh'\right)\text{vol}(\text{AdS}_3)\!\wedge \!J\wedge \!J\wedge\!J\!\wedge \! dr.\nonumber
\end{align}
\end{gather}
As one can see, this expression is proportional up to a boundary term\footnote{Note that using  \eqref{hprofile} and the boundary conditions $h(0)=h(P)=0$ the boundary term does not contribute to the final result for the holographic central charge. As already mentioned, in order to find a finite result the volume of $\text{AdS}_3$ has to be conveniently regularised.}, to our expression \eqref{centralcharge} for the holographic central charge. In 
\cite{Lozano:2020txg,Lozano:2020sae,Lozano:2021rmk} an analogous result was obtained for various $\text{AdS}_2$ backgrounds with $\mathcal{N}=4$ supersymmetries, and it was interpreted as a generalisation of the proposal in 
\cite{Hartman:2008dq}, showing that the central charge in the algebra of symmetry generators of $\text{AdS}_2$ with an electric field is proportional to the square of the electric field. In these references it was shown that the proposal in \cite{Hartman:2008dq} could be extended to fully-fledged $\text{AdS}_2$ string theory set-ups, where it holds for the different branes involved in terms of their electric and magnetic charges. Our result in this section shows that the proposal in \cite{Hartman:2008dq} is not exclusive to $\text{AdS}_2$, but holds in more general settings involving $\text{AdS}$ backgrounds with different dimensionalities. This points at a possible relation to the calculation of the holographic central charge in terms of the regularised on-shell supergravity action. Work is in progress to try to shed more light on this issue.

\section{Conclusions} \label{conclusions}

In this paper we have made progress towards the understanding of $\text{AdS}_3/\text{CFT}_2$ holography with $\mathcal{N}=(0,6)$ supersymmetry. Taking as our starting point the recent local solutions constructed in \cite{Macpherson:2023cbl}, for which the internal space is a foliation of a $\mathbb{CP}^3$ over an interval, and thus bear a close resemblance with the ABJM/ABJ solution, we have proposed a brane set-up and, associated to it, a quiver field theory emerging from the quantisation of the open strings. According to our proposal this field theory should flow in the IR to the 2d CFT dual to the $\text{AdS}_3\times \mathbb{CP}^3$ solutions. We have argued that, in full similarly with the 3d case, supersymmetry should be enhanced from $\mathcal{N}=(0,3)$ for the brane intersection  to $\mathcal{N}=(0,6)$ for the SCFT that emerges in the IR. A difference with the 3d case is that both our solutions and associated field theories preserve half of the supersymmetries of the 3d case. This is consistent with an interpretation of the solutions  as describing $\frac12$-BPS backreacted surface defects within the ABJM/ABJ theory. These defects consist on D8-NS5 branes that reduce the supersymmetries of the ABJ brane set-up by a half and the superconformal algebra to $\mathfrak{osp}(6|2)$. This could represent a conformal embedding of the D4-D8 defects in ABJM proposed in \cite{Fujita:2009kw}, in order to realise edge states in the Fractional Quantum Hall Effect\footnote{Albeit with fractional D4-branes.}. In the absence of D8-branes our proposed brane configuration can also be seen as the intersection of the 
ABJM brane set-up, preserving $\mathcal{N}=3$ in 3d, with the D2-NS5-D4 brane intersection studied in \cite{Kitao:1998mf,Bergman:1999na}, also $\mathcal{N}=3$ supersymmetric in 3d\footnote{See also the closely related brane intersections studied in \cite{Jafferis:2008qz,Imamura:2008nn}, realising Chern-Simons matter theories of quiver type.}. The resulting D2-KK-D6-D4-NS5 intersection would preserve $\mathcal{N}=(0,3)$ supersymmetry in 2d, and these supersymmetries would be left unbroken when the D8-branes are introduced. An alternative interpretation of the $\text{AdS}_3\times \mathbb{CP}^3$ solutions would thus be as holographic duals to the 2d CFTs to which the field theories living in these brane intersections flow in the IR.  Our proposed 2d quivers manifestly realise the embedding of the defect branes within the quiver field theory associated to the ABJM/ABJ theory. It would be interesting to compute observables such as one point correlation functions and the displacement operator \cite{Dibitetto:2017klx}, that further confirm our defect interpretation.  

Interestingly, we have shown that large gauge transformations in the brane set-up induce the generalisation to the massive case of Seiberg duality in ABJM/ABJ theories \cite{Aharony:2008gk,Aharony:2009fc} proposed in \cite{Bergman:2010xd}. This shows that, as in other theories \cite{Benini:2007gx,Aharony:2009fc}, Seiberg duality can be understood geometrically in terms of large gauge transformations. Remarkably, we have obtained the same transformations as the ones proposed in \cite{Bergman:2010xd} for non-supersymmetric 3d Chern-Simons matter theories. This suggests, on the one hand, that the extension of Seiberg duality in ABJM/ABJ to the massive case proposed in \cite{Bergman:2010xd} could hold more generally, and, on the other, that Seiberg duality in the Gaiotto-Tomasiello theories  \cite{Gaiotto:2009mv} ($\mathcal{N}=3$ in 3d) could be realised as a large gauge transformation at the expense of turning one of the field theory directions onto an energy scale, inducing a flow across dimensions to a 2d CFT with $\mathcal{N}=(0,6)$ supersymmetry and an $\text{AdS}_3\times \mathbb{CP}^3$ dual. 

The latter might suggest that $\text{AdS}_4$ solutions to massive Type IIA supergravity with $\mathcal{N}=3$ supersymmetry might not exist beyond the perturbative small mass limit taken in \cite{Gaiotto:2009yz}. We would like to stress that in our solutions the mass parameter can be of the same order as the rest of quantised charges, so the solutions are in this sense fully non-perturbative, and so are their proposed 2d dual CFTs, as opposed to the small deformations of the ABJM/ABJ theory studied in  \cite{Gaiotto:2009mv}. Note as well that the 3d CFTs studied in \cite{Gaiotto:2009mv} were constructed under the assumption that the (small) massive deformations should give rise to different fixed point theories in the same number of dimensions, namely three. 
 It would be interesting to investigate the small mass limit of our solutions and see whether they bear any connection with the perturbative solutions found in \cite{Gaiotto:2009yz}.
 
With respect to the central charge we have made a concrete prediction for the $(0,6)$ SCFTs dual to the $\text{AdS}_3\times \mathbb{CP}^3$ solutions that include sub-leading corrections. However, we have not been able to check this result against a field theory calculation once the rotation of the branes is taken into account and the supersymmetry is reduced to $(0,3)$. As mentioned in the main text, we are unaware of a general result in the literature that relates the level of the superconformal algebra with the R-symmetry anomaly for $(0,3)$ supersymmetry. We hope that our, in this sense, phenomenological, results, stimulate further investigations in this direction. It is likely that the studies in \cite{Franco:2021ixh,Franco:2021vxq} of 2d $(0,1)$ field theories can be successfully used for this purpose. 

More generally, it would be interesting to relate the (holographic) central charge computed in this paper to the exact results for the free energy of the small deformations of the ABJM/ABJ theory with $\mathcal{N}=3$ supersymmetry studied in \cite{Suyama:2010hr,Jafferis:2011zi,Aharony:2010af,Liu:2021njm}, along the lines of our discussion for the massless case.

\section*{Acknowledgements}
We would like to thank Antonio Amariti and Carlos Nunez for useful discussions.
YL and NM are partially supported by the grant
from the Spanish government MCIU-22-PID2021-123021NB-I00. The work of NM is also supported by the Ram\'on y Cajal fellowship RYC2021-033794-I. The work of NP is supported by the Israel Science Foundation (grant No. 741/20) and by the German Research Foundation through a German-Israeli Project Cooperation (DIP) grant ``Holography and the Swampland". The work of AR is partially supported by the INFN grant {\em{Gauge and String Theory (GAST)}} and by the INFN-UNIMIB contract, number 125370/2022.

\end{document}